\documentclass[preprint,aps]{revtex4}
\usepackage{ulem}
\usepackage{amsmath}
\usepackage{color}
\usepackage{graphicx}
\usepackage{epsfig}
\newlength{\upit}\upit=0.1truein

\newcommand{\ltappr}{{{\lower4pt\hbox{$<$} } \atop \widetilde{ \ \ \ }}}
\newcommand{\gtappr}{{{\lower4pt\hbox{$>$} } \atop \widetilde{ \ \ \ }}}
\newlength{\bxwidth}\bxwidth=1.5 truein

\newcommand{\zmatrix}[4]{\left(\begin{matrix}#1 & #2\cr #3&#4\end{matrix}\right)}

\newcommand{\tr}{{\hbox{Tr}}}

\newcommand{\dg}{^{\dagger }}

\newcommand{\up}{\uparrow}
\newcommand{\dw}{\downarrow}

\newlength{\figwidth}
\newlength{\shift}
\newlength{\fight}

\newcommand \bea {\begin{eqnarray} }
\newcommand \eea {\end{eqnarray}}
\newcommand{\bk}{{\bf{k}}}

\newcommand{\bQ}{{\bf{Q}}}
\newcommand{\bR}{{\bf{R}}}
\newcommand{\urs}{URu$_{2}$Si$_{2}$\ }
\newcommand{\ursp}{URu$_{2}$Si$_{2}$}
\newcommand{\Hast}{{\Psi}}
\newcommand{\hast}{{\Psi}}
\newcommand{\psic}{{c}}

\newsavebox{\fmbox}

\begin{document}
\title{Hastatic Order in \urs}

\author{Premala Chandra$^{1}$, Piers Coleman$^{1,3}$, and Rebecca
Flint$^{2}$}

\affiliation{
$^{1}$Center for Materials Theory, Department of Physics and Astronomy,
Rutgers University, 136 Frelinghuysen Rd., Piscataway, NJ 08854-8019, USA}
\affiliation{$^{2}$Department of Physics, Massachusetts Institute for Technology,Massachusetts Avenue, Cambridge, MA 02139-4307,USA}
\affiliation{$^{3}$ Department of Physics, Royal Holloway, University
of London, Egham, Surrey TW20 0EX, UK.}

\date{\today}

\begin{abstract}
\bf
The development of collective long-range order via phase
transitions occurs by the spontaneous breaking of fundamental
symmetries. Magnetism is a consequence of broken time-reversal 
symmetry while superfluidity results from broken gauge
invariance. The
broken symmetry that develops below 17.5K in the heavy fermion compound
\urs has long eluded such identification. Here we
show that the recent observation of Ising quasiparticles in \urs
results from a spinor order parameter 
that breaks {\sl double} time-reversal symmetry, mixing states
of integer and half-integer spin. 
Such ``hastatic order'' hybridizes conduction 
electrons with Ising  $5f^{2}$ states of the uranium
atoms to produce Ising quasiparticles;
it accounts for the large entropy of condensation and the
magnetic anomaly observed in torque magnetometry. Hastatic order
predicts  a tiny transverse moment
in the conduction sea, a collosal Ising anisotropy in the
nonlinear susceptibility anomaly and a resonant
energy-dependent nematicity in the tunneling density of states.
\end{abstract}
\maketitle

%
%
The hidden order (HO) that
develops below 17.5K in the heavy fermion compound \urs 
is 
particularly 
notable, having eluded identification for twenty five
years\cite{Palstra85,Schlabitz86,Mydosh11,Amitsuka94,Haule09,localized,chandra02,Varma,pepin,Morr10,Dubi10,Fujimoto11,Ikeda11}.  Recent
spectroscopic\cite{Santander09,Schmidt10,Aynajian10,Park11,Timusk11},
magnetometric\cite{Okazaki11} and high-field
measurements\cite{Hassinger10,Altarawneh11} 
suggest that the HO is connected with 
the formation of an itinerant heavy electron fluid,
a consequence of 
quasiparticle
hybridization between localized, 
spin-orbit coupled
f-moments and mobile conduction
electrons.
Though the development of 
hybridization at low temperatures is 
usually associated with a crossover, in \urs
both optical \cite{Bonn88} and 
tunnelling \cite{Schmidt10,Aynajian10,Park11}
probes suggest that it develops abruptly at the HO
transition, leading to proposals\cite{Morr10,Dubi10}
that the hybridization is an order parameter.

High temperature bulk susceptibility measurements on \urs show  that the local 5f moments embedded in the
conduction sea are Ising in nature\cite{Palstra85,Ramirez92}, while quantum oscillation (QO) experiments deep within the HO phase 
reveal that the quasiparticles exhibit a 
giant Ising anisotropy\cite{Altarawneh11,Hc2,Altarawneh2}.  The Zeeman splitting depends solely on the c-axis component
of the magnetic field, $\Delta E= g (\theta )\mu_{B}B
$\cite{Altarawneh2}, with a g-factor $g (\theta )=g\cos \theta $, where
$\theta $ is the angle relative to the c-axis. The g-factor anisotropy
exceeds 30, corresponding to an anisotropy of the Pauli susceptibility in excess of
900; this anisotropy is also observed in the angle-dependence of the
Pauli-limited upper critical field of the superconducting
state\cite{Hc2,Altarawneh2}, showing that the Ising quasiparticles pair
to form a heavy fermion superconductor.
This giant anisotropy suggests 
that the f-moment is transferred
to the mobile quasiparticles through hybridization\cite{Osborne02}.
In the tetragonal crystalline environment of \urs such Ising
anisotropy is only natural in integer spin
configurations\cite{Amitsuka94,Flint12}, thus
the most likely valence of the $U$ ions with an integer
spin is a $5f^{2}$ configuration. Moreover,
the observation of paired Ising quasiparticles
in a superconductor with $T_c \sim 1.5 K$ indicates that this $5f^2$
configuration is degenerate to within an energy resolution of
$g\mu_BH_c \sim 5 K$. In \ursp, tetragonal symmetry
protects a magnetic non-Kramers $\Gamma_{5}$ doublet, the candidate
origin of the Ising quasiparticles\cite{Amitsuka94,Ohkawa99}.

The quasiparticle hybridization of half-integer spin 
conduction electrons with an integer
spin doublet in \urs has profound implications for 
hidden order; such 
mixing can not occur without broken
{\sl double} time-reversal symmetry. 
Time-reversal,   $\hat \Theta $,
is an  anti-unitary quantum operator with {\sl no} associated 
quantum number\cite{sakurai}. However double-reversal
$\hat  \Theta^{2}$,  equivalent to a $2\pi$ rotation, 
forms a unitary operator with an associated
quantum number, the ``Kramers index'' $K$\cite{sakurai}.  The
Kramers index, $K= (-1)^{2J}$ of a
quantum state of total angular momentum $J$ defines the phase factor
acquired by its wavefunction after two successive time-reversals,
$\hat  \Theta^{2}\vert \psi \rangle = K \vert \psi \rangle
= \vert \psi^{2\pi}\rangle 
 $.  An integer
spin state $\vert \alpha \rangle$ is unchanged by a $2\pi$ rotation, so
$\vert \alpha ^{2\pi}\rangle = +\vert \alpha \rangle $ and $K=1$. 
However, conduction electrons with half-integer spin states, $\vert \bk \sigma \rangle $, where
$\bk$ is momentum and $\sigma$ is the spin component, change sign,
$\vert \bk \sigma^{2\pi}\rangle = - \vert \bk \sigma \rangle $, so
$K=-1$.

While conventional magnetism breaks time-reversal symmetry,
it is invariant under $\hat  \Theta^{2}$ so the 
Kramers index is conserved. 
However
in \ursp, the hybridization 
between integer and half-integer spin states
requires a  quasiparticle mixing term of the form 
${\cal  H}= (
\vert \bk \sigma \rangle V_{\sigma \alpha } (\bk )
\langle \alpha  \vert + {\rm H.c})$ 
in the low energy fixed point Hamiltonian.
After two successive time-reversals
\begin{equation}\label{}
 \vert \bk \sigma \rangle V_{\sigma \alpha } (\bk )\langle \alpha
 \vert\rightarrow 
\vert \bk \sigma ^{2\pi}\rangle V^{{2\pi}}_{\sigma \alpha } (\bk )
\langle \alpha ^{2\pi}\vert
= - \vert \bk \sigma \rangle V^{2\pi}_{\sigma \alpha } (\bk )\langle \alpha  \vert.
\end{equation}
Since the microscopic Hamiltonian is time-reversal invariant, 
it follows that $V_{\sigma \alpha } (\bk ) =-V_{\sigma \alpha }^{2\pi} (\bk )$; 
the hybridization thus breaks
time-reversal symmetry in a fundamentally new way, 
forming an order parameter 
that, like a spinor, reverses under $2\pi$ rotations.  
The resulting 
 ``hastatic (Latin: {\sl spear}) order'', is a state of matter that
breaks both single and double time-reversal symmetry and
 is thus distinct from conventional magnetism.

Indirect support for time-reversal symmetry-breaking in the hidden
order phase of \urs is provided by 
recent magnetometry measurements that 
indicate the development of an anisotropic
basal-plane spin susceptibility, 
$\chi_{xy}$, at the hidden order
transition\cite{Okazaki11}. 
$\chi_{xy}$ 
is a conduction electron response to scattering off the hidden
order (c.f. Fig. 2.), leading to a scattering 
matrix of the form 
$t(\bk) = (\sigma_x + \sigma_y) d(\bk)$,
where $d(\bk)$ is the scattering amplitude. 
$t (\bk )$ has been linked to a  
spin nematic state\cite{Fujimoto11}, under the special condition that 
$d (-\bk )= -d^{*} (\bk )$ to avoid time-reversal symmetry breaking.
However, if the scattering process involves resonant 
hybridization in the f-channel, then $d (\bk )$ is associated
with resonant scattering off the f-state, a process with 
a real, even parity scattering amplitude, $d(\bk) =
d(-\bk)$. In this case, the observed t-matrix is necessarily 
odd under time-reversal in the hidden order phase.

Hybridization
in heavy fermion compounds 
is usually driven by valence fluctuations mixing
a
ground-state Kramers doublet and an excited singlet (cf. Fig. 1a). 
In this case, the hybridization
amplitude is a scalar that develops 
via a crossover, leading to mobile heavy fermions.
However valence fluctuations from a $5f^{2}$ ground-state create
excited states with an odd number of electrons and hence a Kramers
degeneracy (cf. Fig. 1b). 
Then the quasiparticle hybridization 
has two components, $\hast_{\sigma }$,
that determine the mixing of the excited Kramers doublet into the
ground-state.  These two amplitudes form a spinor defining the 
``hastatic'' 
order parameter
\begin{equation}\label{}
\Hast= \left(\begin{matrix} \hast_{\up }\cr \hast_{\dw}\end{matrix}
\right).
\end{equation}
The presence of distinct up/down hybridization
components indicates that $\Psi$ 
carries the global spin quantum number;
its development must 
now break double time-reversal and spin rotational
invariance via a phase transition.

Under pressure, \urs  undergoes a first-order phase transition from
the hidden order (HO) state to an antiferromagnet (AFM) 
\cite{Amitsuka07}.  These two states are remarkably close in energy
and share
many key features\cite{Jo07,Villaume08,Hassinger10} including
common Fermi surface pockets; this motivated
the recent proposal that despite the first order transition
separating the two phases, they are linked by ``adiabatic 
continuity,''\cite{Jo07} corresponding to a notional rotation
of the HO in internal parameter space \cite{Haule09,Haule10}.
In the magnetic phase, this spinor points along the c-axis
\begin{equation}\label{}
\Hast_A\sim \left(\begin{matrix} 1 
\cr 0 
\end{matrix} \right)  
\hbox{,  }
\Hast_B\sim 
 \left(\begin{matrix} 0
\cr 1 
\end{matrix} \right)  
\end{equation}
corresponding to time-reversed configurations on alternating layers A
and B, leading to a large staggered Ising moment.
For the HO state, 
the spinor points in the basal plane
\begin{equation}\label{}
\Hast_A \sim  \frac{1}{\sqrt{2}}\left(\begin{matrix} 
e^{-i\phi/2} \cr  e^{i \phi/2 } \end{matrix} \right)
\hbox{,  }
\Hast_B \sim  \frac{1}{\sqrt{2}}\left(\begin{matrix} 
- e^{-i\phi/2} \cr  e^{i \phi/2 } \end{matrix} \right),
\end{equation}
where again, $\Psi_{B}= \Theta \Psi_{A}$ and it
is protected from developing a 
large moment by the pure Ising character of the $5f^{2}$ ground-state. 
%
%
%

Hastatic order permits a direct realization of the adiabatic
continuity between the HO and AFM 
 in terms of 
a single Landau functional for the free energy
\begin{equation}\label{}
f[T,P,B_{z}] = [\alpha (T_{c}-T)-\eta_z B_{z}^{2}]\vert \Psi
\vert^{2}+ \beta \vert \Psi \vert^{4} - \gamma (\Psi \dg \sigma_{z}\Psi )^{2}
\end{equation}
where $\gamma = \delta
(P-P_{c})$ is a pressure-tuned anisotropy term. The unique
feature of the theory is that the non-Kramers doublet has Ising
character, and only couples to the z-component of the magnetic
field $B_{z}=B \cos\theta$. The resulting Ising splitting of the
non-Kramer's doublet suppresses the Kondo effect, giving rise to 
the $B_{z}^{2}$ term in the quadratic coefficient, where the coefficient
$\eta_z$ is of order $1/T_{HO}^2$. 
\cite{Supplementary}
The phase diagram predicted by this free energy is shown in Fig. 1 (c).
When $P<P_{c}
(T)$, the vector $\Hast\dg  \vec{\sigma}\Hast = \vert \Psi
\vert^{2} (n_{x},n_{y},0)
$ lies in the basal plane, resulting in hastatic order.
At $P=P_{c}$, there is a first order ``spin-flop'' into an magnetic
state where $\Hast\dg  \vec{\sigma}\Hast = \vert \Psi\vert^{2}(0,0,\pm 1)$
lies along the c-axis. 

Adiabatic continuity provides a natural interpretation of 
the soft longitudinal 
spin fluctuations observed to develop 
in the HO state \cite{Broholm91}
as an incipient Goldstone excitation between the two 
phases \cite{Haule10}.  
In the HO state, 
rotations between hastatic and AFM order will lead
to a gapped Ising collective mode 
which we identify with 
the longitudinal spin fluctuations observed in inelastic neutron
scattering \cite{Broholm91}.
At the first order line, $P=P_{c}$, the quartic anisotropy term vanishes;
we predict
that the gap to longitudinal spin
fluctuations will soften according to $\Delta \propto \sqrt{\gamma|\Hast|^{2}}\sim |\Hast |\sqrt{P_{c} (T) -P}$
(see Supplementary material).
Experimental observation of this feature would provide
confirmation of the common origin of the hidden and AFM order. 


Another prediction of the phenomenological theory is 
the development of a nonlinear susceptibility anomaly with a collosal Ising
anisotropy.
 From the Landau theory\cite{Supplementary}, the jump in
the specific heat $\Delta C$, the susceptibility and nonlinear
susceptibility anomalies {$d\chi_{1}/dT$ and $\Delta\chi_{3}$} obey the relation
$\frac{\Delta C}{T_{c}} \Delta\chi_{3}= 12 \left(\frac{ d \chi_{1}}{dT}
\right)^{2}$, where $\frac{d\chi_{1}}{dT} = - \frac{\alpha }{2 \beta
}\eta_z\cos^{2 } \theta $, so that $\chi_{3}\propto \cos^4 \theta $.
A large anomaly in the c-axis nonlinear susceptibility of \urs has been
observed at $T_{c}$\cite{Miyako91,Ramirez92}, but its Ising
anisotropy has never been quantified. 
The predicted development of a collosal Ising anisotropy in the
zero-field non-linear susceptibility at the hidden order transition is
another consequence of hastatic order.

We now present a model that relates hastatic order to the valence fluctuations 
in \ursp,
based on a two-channel Anderson lattice model.
The uranium ground-state 
is a $5f^2$ Ising $\Gamma_5$ doublet\cite{Amitsuka94},
$\vert \pm  \rangle = a  \vert \pm 3 \rangle +b \vert \mp
1\rangle$, written in terms of $J = 5/2$ f-electrons 
in the three 
tetragonal orbitals $\Gamma_7^{\pm}$ and $\Gamma_6$
\begin{eqnarray}\label{nk2}
\vert +\rangle&=& (a f\dg_{\Gamma_{7}^{-}\dw
}f\dg_{\Gamma_{7}^{+}\dw }
+ b f\dg_{\Gamma_{6}\up}f\dg_{\Gamma_{7}^{+}\up})\vert 0 \rangle
\cr
\vert -\rangle&=& (a f\dg_{\Gamma_{7}^{-}\up
}f\dg_{\Gamma_{7}^{+}\up }
+ b f\dg_{\Gamma_{6}\dw}f\dg_{\Gamma_{7}^{+}\dw})\vert 0 \rangle.
\end{eqnarray}
The lowest lying
excited state is most likely the $5f^3$ ($J=9/2$) state, but for simplicity here
we take it to be 
the symmetry-equivalent $5f^{1}$ state.
Valence fluctuations from the ground state
({5f$^2$} $\Gamma_5$) to the excited state ({5f$^1$} $\Gamma_7^+$) occur in two orthogonal
conduction
channels,\cite{Cox96,CoxZawad} $\Gamma_7^-$ and $\Gamma_6$.
This allows us to read off
the hybridization matrix elements of the Anderson model
\begin{equation}
\label{VF1} H_{VF}(j) = V_6 \psic_{\Gamma_6 \pm}\dg(j) |\Gamma_7^+
\pm\rangle \langle \Gamma_5 \pm| + V_7 \psic_{\Gamma_7 \mp}\dg(j)
|\Gamma_7^+ \mp \rangle\langle \Gamma_5 \pm| + \mathrm{H.c.}.
\end{equation}
where $\pm$ denotes the ``up'' and ``down''
states of the coupled Kramers and non-Kramers doublets. The field
$\psic\dg_{\Gamma \sigma}(j) = \sum_\bk
\left[\Phi\dg_\Gamma(\bk)\right]_{\sigma \tau} c\dg_{\bk\tau}
\mathrm{e}^{-i\bk\cdot \bR_j}$ creates a conduction electron at site
$j$ with spin $\sigma$ in a Wannier orbital with symmetry $\Gamma\in
\{6,7\}$, while $V_6$ and $V_7$ are the corresponding hybridization
strengths. 
The full model is then written 
\begin{equation}\label{}
H = \sum_{ \bk \sigma }\epsilon_{\bk }c\dg_{\bk \sigma }c_{\bk \sigma}
+ \sum_{j}\left[H_{VF} (j)+ H_{a} (j) \right]
\end{equation}
while  $H_a(j) = \Delta E \sum_\pm|\Gamma_7 \pm,j\rangle \langle \Gamma_7 \pm,j| $
is the atomic Hamiltonian.

Hastatic order is revealed by factorizing the Hubbard
operators, 
$|\Gamma_7^+ \sigma\rangle\langle \Gamma_5 \alpha| = \hat \hast\dg_\sigma \chi _\alpha.
$
Here $|\Gamma_5 \alpha\rangle = \chi\dg_\alpha
|\Omega\rangle$ is the non-Kramers doublet, represented by the
pseudo-fermion
$\chi \dg_{\alpha }$, while
$\hat \hast_\sigma\dg$ is a {slave} boson{\cite{Coleman83}} representing
the excited $f^{1}$ doublet
$|\Gamma_7^+ \sigma\rangle
=\hat\hast\dg_\sigma|\Omega\rangle$.
Hastatic order is the condensation of this boson $
\hast_{\sigma }\dg\chi_{\alpha} \rightarrow \langle\hat  \hast_{\sigma }\rangle 
\chi_{\alpha }$, generating a
hybridization between the conduction electrons and
the Ising $5f^{2}$ state while also breaking double time reversal.  The $\Gamma_5$
doublet has both magnetic and quadrupolar moments
represented by
 $\chi \dg \vec{\sigma} \chi=
 (\mathcal{O}_{x^{2}-y^{2}},\mathcal{O}_{xy},m^{z}) $, where 
$m^z$ is the Ising magnetic moment and $\mathcal{O}_{x^2-y^2}$ and
$\mathcal{O}_{xy}$ are quadrupole moments.
The tensor 
product $Q_{\alpha \beta } \equiv \Psi_{\alpha }\Psi \dg_{\beta }$
describes the development of composite order between 
the non-Kramers doublet and the spin density of 
conduction electrons. Composite order has been 
considered by several earlier authors
in the context of 
two channel Kondo lattices\cite{Cox96,CATK,Hoshino11} in which the
valence fluctuations have been integrated out.
However, by factorizing the composite order in terms 
of the spinor  $\Psi_{\alpha }$, we are able to directly understand
the development of coherent Ising quasiparticles and the broken double
time-reversal.

Using this factorization, we can rewrite the valence fluctuation term as,
\bea
\label{VF3}
H_{VF}(j) = \sum_\bk c\dg_{\bk\sigma} \hat{\mathcal{V}}_{\sigma
\eta}(\bk,j) \chi _\eta(j) \mathrm{e}^{-i\bk\cdot \bR_j} +
\mathrm{H.c.}
\eea
where
$
\hat{\mathcal{V}}(\bk,j) = V_6 \Phi_{\Gamma_6}{\dg}(\bk)
\hat B\dg _j +   V_7 \Phi_{\Gamma_7^-}{\dg}(\bk)\hat B\dg_j\sigma_1, \quad \hat B_{j} =
\zmatrix{\hat \hast
_\up}{0}{0}{\hat \hast_\dw}.
$
In the ordered state, 
$B_{j}= \langle \hat
B_{j}\rangle $ is replaced by its expectation value, so that in the
HO state
\begin{equation}
\langle B\dg_j\rangle = |\Hast|\zmatrix{\mathrm{e}^{i(\bQ \cdot \bR_j + \phi)/2}}{0}{0}{\mathrm{e}^{-i(\bQ \cdot \bR_j + \phi)/2}} \equiv |\Hast| U_j,
\end{equation}
with magnitude $|\hast|$.  The internal angle, $\phi$, rotates $B_j$
within the basal plane. 

As the HO and AFM appear to share a single commensurate wavevector, $\bQ =
(0,0,\frac{2\pi}{c})$\cite{Jo07,Villaume08,Hassinger10}, we use this
wavevector here.  
It is convenient to
absorb the unitary matrix, $U_j$ into the pseudo-fermion, $\tilde{\chi
}_j = U_j \chi _j$.  {This gauge transformation transfers the charge from the slave boson to the pseudo-fermion, making it a charged quasiparticle}.  In this gauge, one
channel ($\Gamma_6$) is uniform, while the other ($\Gamma_7^-$) is
staggered 
\bea 
H_{VF} = \sum_\bk c\dg_\bk
\mathcal{V}_6(\bk) \chi _\bk + c\dg_\bk \mathcal{V}_7(\bk) \chi _{\bk+\bQ} +
\mathrm{h.c.}  
\eea
where the hybridization form factors $
\mathcal{V}_7(\bk) = V_7 \Phi_7{\dg}(\bk) \sigma_1$ and $\mathcal{V}_{6}
(\bk )= V_{6}\Phi_{6}{\dg} (\bk )$.

There are two general aspects of this condensation that deserve
special comment. First, the two-channel Anderson impurity model is
known to possess a non-Fermi liquid ground-state 
with an entanglement entropy of $\frac{1}{2} k_{B}\ln  2$\cite{Bolech02}. 
The development of hastatic order liberates
this zero-point entropy, accounting naturally for the large entropy
of condensation.
As a slave boson, $\Psi $ carries both
the charge $e$ of the electrons and the local gauge charge $Q_{j}=
\Psi\dg _{j}\Psi_{j}+\chi \dg_{j}\chi_{j}$ of constrained valence fluctuations,
its condensation gives a mass to their relative
phase via the Higg's mechanism\cite{marston03}.  But as a Schwinger
boson,  the condensation of $\Psi $
this process breaks the SU (2) spin symmetry. 
In this way
the hastatic boson can be regarded as a magnetic analog of the Higg's boson.


One of the key elements of the hastatic theory is the
formation of mobile Ising quasiparticles, and the observed Ising
anisotropy enables us to set some of the parameters of the theory. 
The full anisotropic g-factor is a combination of f-electron
and  conduction electron components given by 
$g (\theta )\approx g^{*}\cos \theta + g_{c}\left(\frac{T_{K}}{D}
\right)$ where $g_{c}=2$ and the factor $T_{K}/D$ derives from the
small conduction electron admixture  in the quasiparticles.
 Experimentally\cite{Altarawneh11}, the
g-factor anisotropy is in excess of $30$, i.e. $g_{z}/g_{\perp }\sim D/T_{K} \gtappr 30$,
which enables us to phenomenologically set a lower bound on 
$D/T_{K}$ in our model. 
The g-factor is defined in terms of the Zeeman splitting 
of the heavy fermion dispersion, 
$\Delta E_{\bk  \eta } =
\vert  E_{\bk  \eta \up} - E_{\bk \eta  \dw }\vert  = g_{\bk \eta
}(\theta )\mu_B B$, where $\eta \in [1,4]$  is a band-index. 
Fig.  3(a) shows the Fermi
surface averaged g-factor, defined by
\begin{equation}\label{}
\overline{g (\theta )} = \frac{
\sum_{\bk \eta } g_{\bk \eta } (\theta ) \delta (E_{\bk\eta })}
{\sum_{\bk \eta } \delta (E_{\bk\eta })}
\end{equation}
%
calculated within the mean-field hastatic model, 
as a function of field angle to the c-axis, choosing the lower-bound 
estimate $D/T_{K}\sim 30$. 

Another key aspect of the hastatic picture is that there must be time-reversal symmetry breaking in \emph{both} the
HO and AFM phases, manifested
by a staggered moment; in the
AFM this leads to   a large c-axis f-electron moment, but in the
HO it becomes  a small transverse moment carried by 
conduction electrons, $\vec{m}_c = -g \mu_B \tr \vec{\sigma}
\mathcal{G}^c(\bk,\bk+\bQ)$, where ${\cal G}^{c}$ is the 
conduction electron Green's function\cite{Supplementary}.  
The small magnitude of the induced  moment
is a consequence of the Clogston-Anderson compensation
theorem that states that changes in the conduction electron
magnetization are set by the same ratio $T_{K}/D$ that determines
the g-factor anisotropy\cite{Anderson61}.
There will also be a small mixed-valent contribution 
from the excited Kramers doublet, $\vec{m}_1 \propto \langle \Hast\dg \vec{\sigma} \Hast \rangle$. 
The angle of the moments in the plane is controlled by the internal hastatic angle, $\phi$.
Fig. (\ref{mag_chi}b) shows the temperature dependence of the 
magnetic moment calculated 
for a case where $D/T_{K} \approx 30$, 
 for which $m_\perp(0)$ =0.015$\mu_{B}$, which is an upper bound
for the predicted conduction electron moment.
Neutron scattering measurements on \urs have 
placed bounds on the c-axis magnetization of the f-electrons 
using a momentum transfer $Q$ in the basal plane.
Detection of an $m_\perp(0)$ carried by conduction electrons, 
with a small scattering form-factor,
will require
high-resolution measurements with a c-axis momentum transfer.
We note that there have been reports from 
$\mu$SR and NMR measurements\cite{Amitsuka03,Bernal04} 
of very small intrinsic basal plane fields in \ursp, 
which  are consistent with this theory.

Although the conduction electrons develop a magnetic moment, the non-Kramers
5f$^{2}$ state carries no dipolar or quadrupolar moment, $\langle \chi \dg \vec{\sigma }\chi \rangle  = 0$.
In the microscopic model the quadrupolar moments vanish
because of the d-wave form factor between the $\Gamma_{6}$ and
$\Gamma_{7^{-}}$ channels\cite{Supplementary}. 
The absence of a charge quadrupole implies there will be no 
lattice distortion associated with hastatic order.
By contrast, hastatic order does induce a weak broken 
tetragonal symmetry in the spin channel.
In the HO state,
the inter-channel components of the hastatic t-matrix,
$\hat{\mathcal{V}}_6 \hat{\mathcal{V}}_7\dg \propto \sigma_x
+\sigma_y$, break tetragonal symmetry in the spin channel, resulting
in a nonzero spin susceptibility within the conduction fluid,
\begin{equation}
\chi_{xy} = -(g\mu_B)^2\tr \sigma_x \mathcal{G}^c(\bk,\bk+\bQ) \sigma_y \mathcal{G}^c(\bk+\bQ,\bk) 
\propto (\tr\hat{\mathcal{V}}_6 \hat{\mathcal{V}}_7\dg)^2
\end{equation}
of a magnitude of order $\left(
\frac{T_{K}}{D} \right)^{2}$, 
that onsets at the hidden order temperature as $|\Hast|^4 \sim
(T_c-T)^2$ as shown in Fig. (3c).

Hastatic order also manifests itself as an anisotropic
hybridization gap, which vanishes along lines in momentum space,
giving rise to a ``V''-shaped {density of states due to the} partial gapping of the Fermi surface, as
shown in Fig (4){, which will be smeared out by disorder in the real material.}
The
anisotropy of the hybridization also breaks tetragonal symmetry,
giving rise to a energy
dependent nematicity $\eta (E)$
that peaks over a narrow energy window  
around the f-resonance.  Some of this nematicity is
present at the Fermi surface, accounting for the splitting seen in
quantum oscillation frequencies\cite{Ohkuni99} and cyclotron resonance
experiments\cite{Matsuda11}.  
An ideal way to verify this
prediction is via scanning tunneling spectroscopy  
(STS), where the measured differential conductivity,
$dI/dV\propto A (eV,\vec{x})$ 
is proportional to the local density of states $A (eV,\vec{x})$ at
position $\vec{x}$  on the surface. 
A measure of the broken tetragonal symmetry is provided by the 
``nematicity''
\begin{equation}
\eta (V) = \frac{
\overline{\ \frac{dI}{dV} (x,y)
\ {\rm sign} (x\ y)}}{\left(
\overline{ 
\left(
\frac{dI}{dV} \right)^{2}} 
-\left( \overline{ \frac{dI}{dV} }\right)^{2}
 \right)^{\frac{1}{2}}}.
\end{equation}
Here 
$(x,y)$ is the co-ordinate relative to the center of the unit
cell and the overbar denotes an average over the unit cell.
The resonant scattering of the hastatic order causes this quantity
to vary rapidly as a function of voltage, over an energy range of order
the Kondo temperature $T_{K}$. Fig.  
(\ref{didvfig}) shows the variation of the nematicity, calculated
within our model of hastatic order. As one passes through
the Kondo resonance, this quantity is found to change sign, and to
peak at a maximum value of  approximately 50\%.

So far we have focused on the mean field consequences of hastatic
order.  The recently observed softening of the commensurate longitudinal spin
fluctuations at $T_c$\cite{Niklowitz11} suggests that hastatic order
melts via phase fluctuations.  While the hybridization spinor itself,
$\langle \Hast \rangle$, can only become nonzero below the phase
transition at $T_c$, we expect that its amplitude, $\langle \Hast\dg
\Hast\rangle$, will persist to higher temperatures.  As $\langle
\Hast\dg \vec{\sigma}\Hast\rangle$ remains zero above $T_c$, only the
non-symmetry breaking (uniform, intrachannel) components of the
hybridization can develop: $\hat{\mathcal{V}}_6
\hat{\mathcal{V}}_6\dg$ and
$\hat{\mathcal{V}}_7\hat{\mathcal{V}}_7\dg$ will emerge via a
crossover at a higher temperature $T^*$ to create an incoherent Fermi
liquid,
consistent with the heavy mass inferred from thermodynamic and optical
measurements\cite{Palstra85,Timusk11}
and the development of Fano signatures in both 
scanning and point-contact tunneling
spectroscopy\cite{Schmidt10,Aynajian10,Park11}. The symmetry-breaking,
interchannel components, $\hat{\mathcal{V}}_6 \sigma_1
\hat{\mathcal{V}}_7\dg$ will always develop precisely at the HO
transition.
Another aspect of
experiments that is not covered by our mean field description, is the
observation of gapped, low-energy incommensurate fluctuations around a
Q-vector $Q^*= (1\pm 0.4,0,0)$ in the
hidden order phase, which appears to be a sign of an unfulfilled 
predisposition towards an incommensurate phase, probably driven by
partial Fermi surface nesting.  These effects 
lie beyond a mean-field description,
but would emerge from the Gaussian fluctuations about the
mean-field theory .

Though we have discussed hastatic order in the context of \ursp, it should be a more
a widespread phenomenon associated with hybridization in any
f-electron material whose unfilled f-shell contains a
geometrically-stabilized non-Kramers doublet
As such, we expect realizations of hastatic order in other 5f uranium and 4f praseodymium intermetallics.



\begin{itemize}
\item[]{\bf Acknowledgments}  An early version of this work was begun in
collaboration with Patrick Fazekas, since deceased. We thank
N. Andrei, S. Burdin, B. Coleman, 
N. Harrison, K. Haule, G. Kotliar, P. Lee, G. Luke, L. Greene, Y. Matsuda,
J. Mydosh, P. Niklowitz, C. P\' epin, T. Senthil, A. Toth and
T. Timusk for useful discussions.  We acknowledge funding from the
Simons Foundation (Flint), the National Science Foundation grants DMR
0907179 (Flint, Coleman) and grant 1066293 (Flint, Chandra, Coleman)
while at the Aspen Center for Physics. We are grateful for the
hospitality of the Aspen Center for Physics.

 \item[]{\bf Competing Interests} The authors declare that they have no
competing financial interests.

 \item[]{\bf Correspondence} and requests for materials
should be addressed to Piers Coleman.~(email: coleman@physics.rutgers.edu).
\end{itemize}
\newpage

\noindent {\bf Figure Captions}

\renewcommand{\labelenumi}{{\bf Fig.} \theenumi .}
\begin{enumerate}

\item (a) A normal Kondo
effect occurs in ions with an odd number of f-electrons, where the
ground state is guaranteed to be doubly degenerate by time-reversal
symmetry (known as a Kramers doublet).  Virtual valence fluctations to
an excited singlet state are associated with a scalar hybridization.
(b) In  \ursp, quasiparticles inherit an Ising symmetry from a 
$5f^{2}$ non-Kramers doublet.  Loss or gain of an electron
necessarily leads to an excited Kramers doublet, and the development
of a coherent hybridization is associated with a two-component spinor
hybridization that carries a \emph{magnetic} quantum number and must
therefore develop at a phase transition. 
(c) Phase diagram for hastatic order, showing how tuning the parameter
$\lambda\propto (P-P_c)$.
leads to a spin flop between hastatic order
and Ising magnetic order. Inset: at the 1st order line, the longitudinal spin
gap is predicted to vanish as 
$\Delta \propto \sqrt{P_c - P}$.
(d) Polar plot showing the predicted $\cos^4 \theta$ form of the non-linear              
susceptibility $\chi_3$ induced by hastatic order, where $\theta$ is the angle of the    
field from the c-axis.  
\label{doublets}

\item \label{hastafig1}Phenomenological interpretation of
the anomalous spin susceptibility in \ursp. (a) The anomalous {spin}
susceptibility is given by conduction electrons scattering off the
hidden order, sandwiched between $\sigma_x$ and $\sigma_y$
vertices. (b) The anomalous scattering t-matrix can be rewritten 
as a resonant scattering off the order parameter.

\item \label{mag_chi}
Magnetic response of hastatic order.  
(a) Polar plot of calculated g-factor, $g (\theta )$ averaged over the Fermi
surface, as a function of magnetic field angle $\theta$ (see SOM for
details), compared with
results of Altarawneh et al. \cite{Altarawneh11}, overlaid in green.
(b) As a consequence of the broken
time-reversal symmetry, we predict a staggered conduction electron
moment that onsets at the HO transition with a linear
$T_c-T$ temperature dependence (staggering pattern shown in inset).
The magnitude of this moment is governed by $T_K/D \sim .01\mu_B/U$.
(c)
We have calculated the tetragonal symmetry breaking component of the
uniform susceptibility, $\chi_{xy}(T)$.  To compare our
results to Okazaki \emph{et al}\cite{Okazaki11} (overlaid as green squares), we have plotted the two-fold oscillation amplitude of the
magnetic torque, $A$ (in black), where $A \cos 2\phi \equiv
\frac{\tau_{2\phi}}{V} = -\frac{\mu_0 H^2}{V} \cos 2\phi \chi_{xy}(T)$.  $A$ goes as $(T_c-T)^2$ just below the HO
transition.  For details of our calculation, including parameter
choices, please see the supporting online
material\cite{Supplementary}.

\item \label{didvfig}
Density of states and resonant nematicity predicted by theory. 
Upper panel: density of states as a function of energy
predicted by model calculation
(blue line), showing f and conduction electron components.  Red line, 
voltage dependence of nematicity $\eta (V)$ in model calculation of
scanning tunneling spectrum. 
Lower panels: spatial dependence of density of states for
selected bias voltages in model calculation of scanning tunneling
spectrum, showing the resonant character of the
nematicity.

\end{enumerate}
\vfill \eject 
\label{}
\setcounter{figure}{0}
\newpage
\begin{figure}[t]
\centerline {\includegraphics[width=1.3\textwidth]{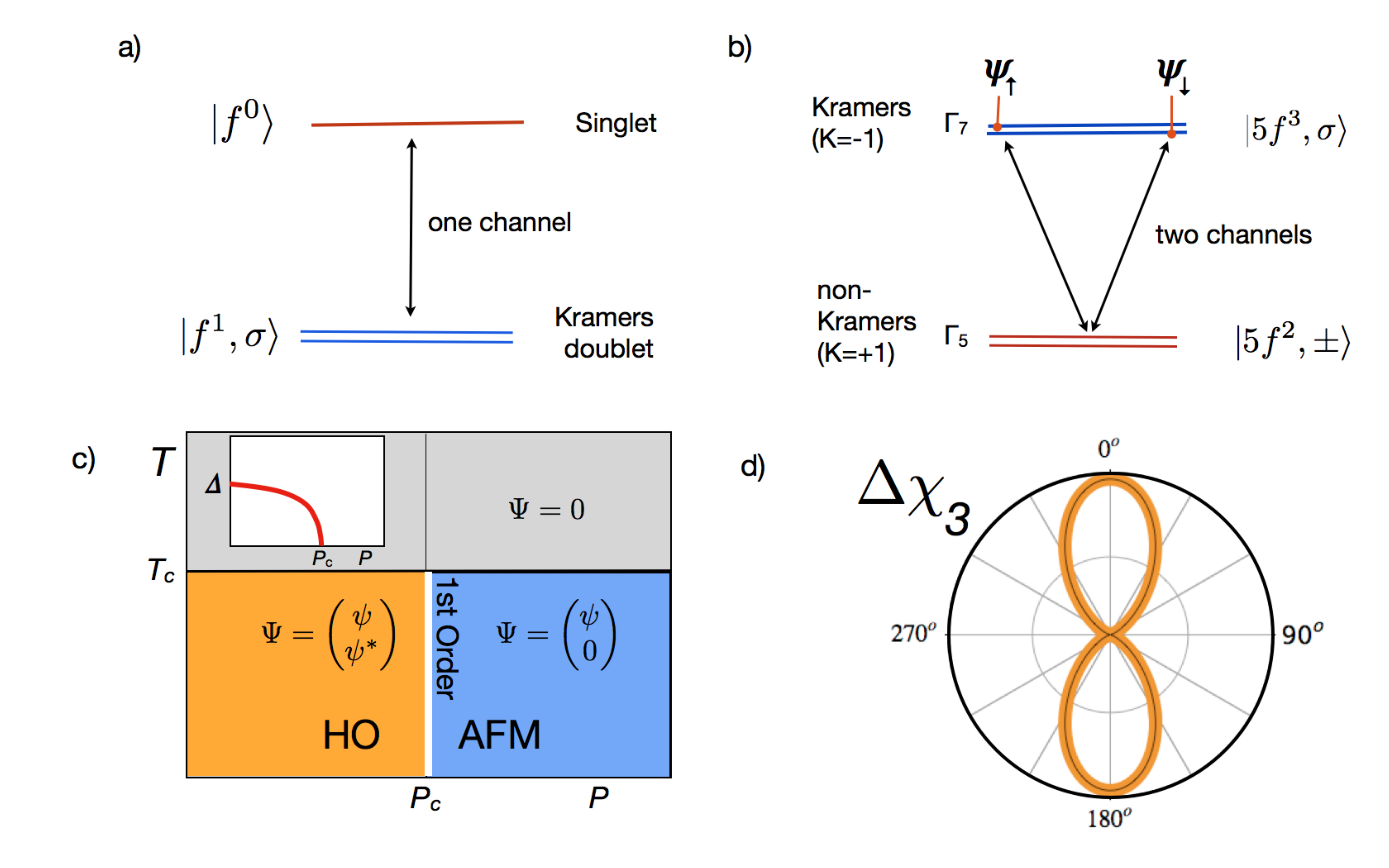}}
{\vskip 5cm\Large \sffamily Figure 1.}
\end{figure}
\newpage
\begin{figure}[t]
\centerline{\includegraphics[width=7in]{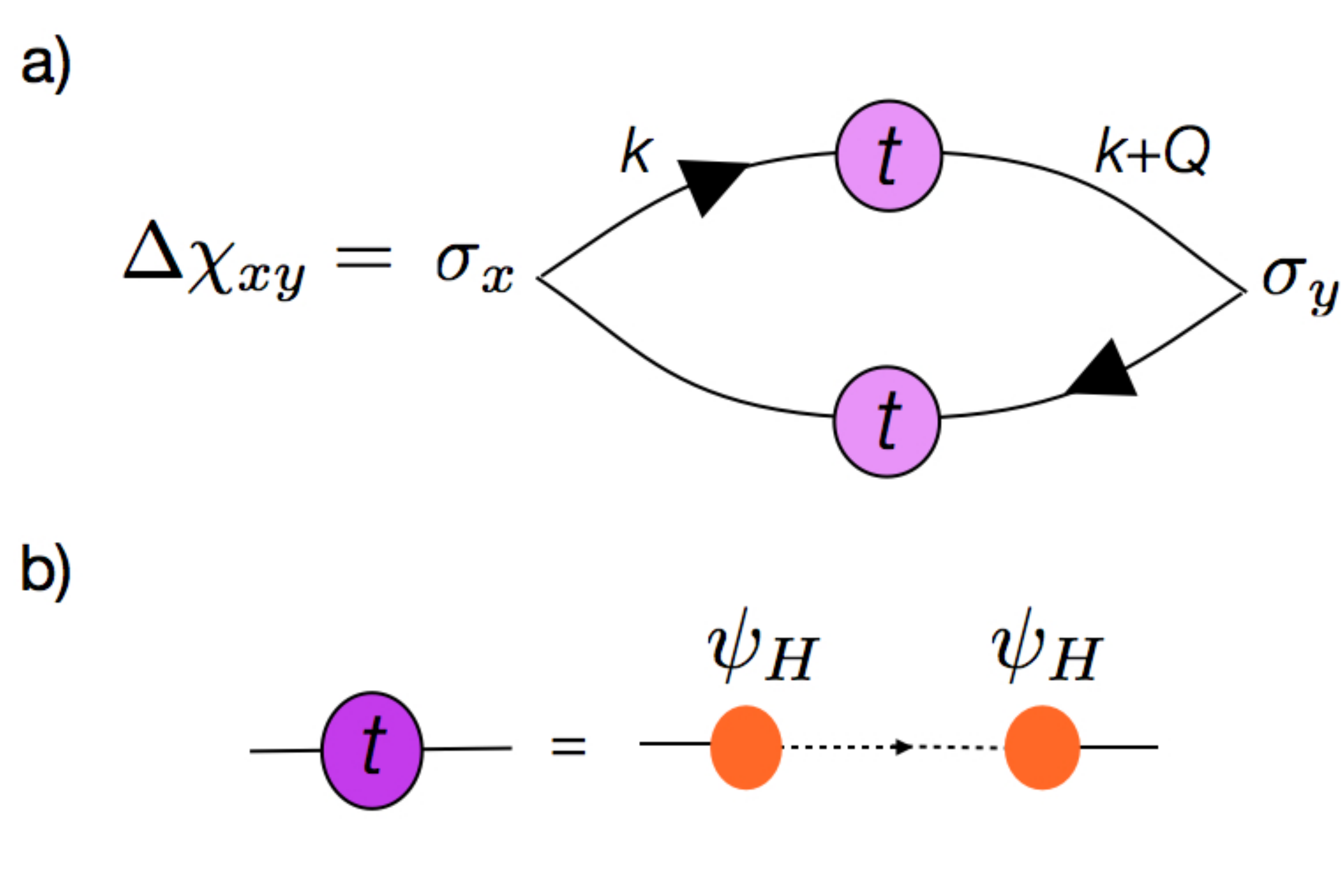}}
{\vskip 5cm\Large \sffamily Figure 2.}
\end{figure}
\newpage
\begin{figure}[t]
\centerline{\hskip 2cm\includegraphics[width=1.2\textwidth]{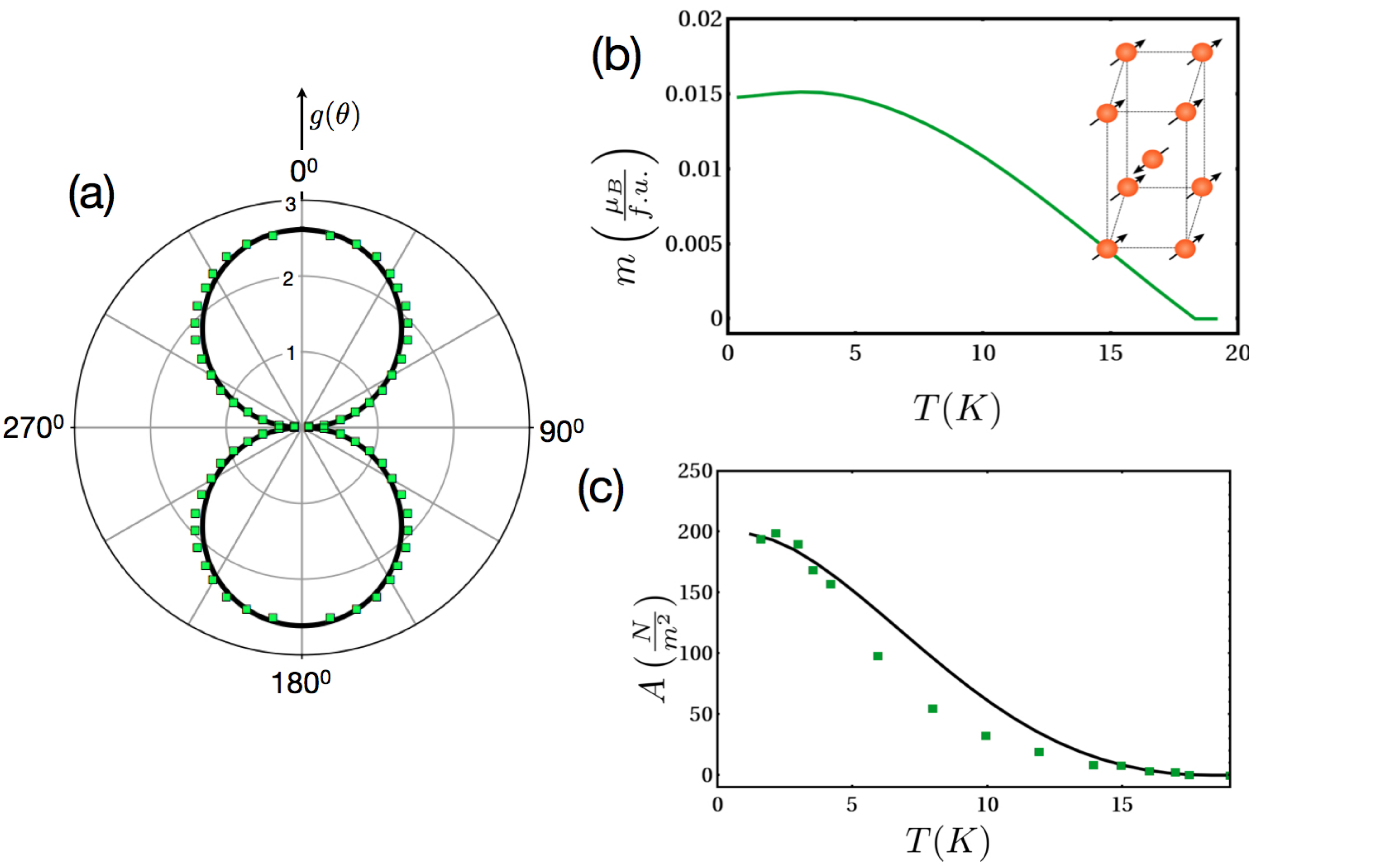}}
{\vskip 5cm\Large \sffamily Figure 3.}
\end{figure}
\newpage
\begin{figure}[t]
\centerline{\hskip 1cm\includegraphics[width=6in]{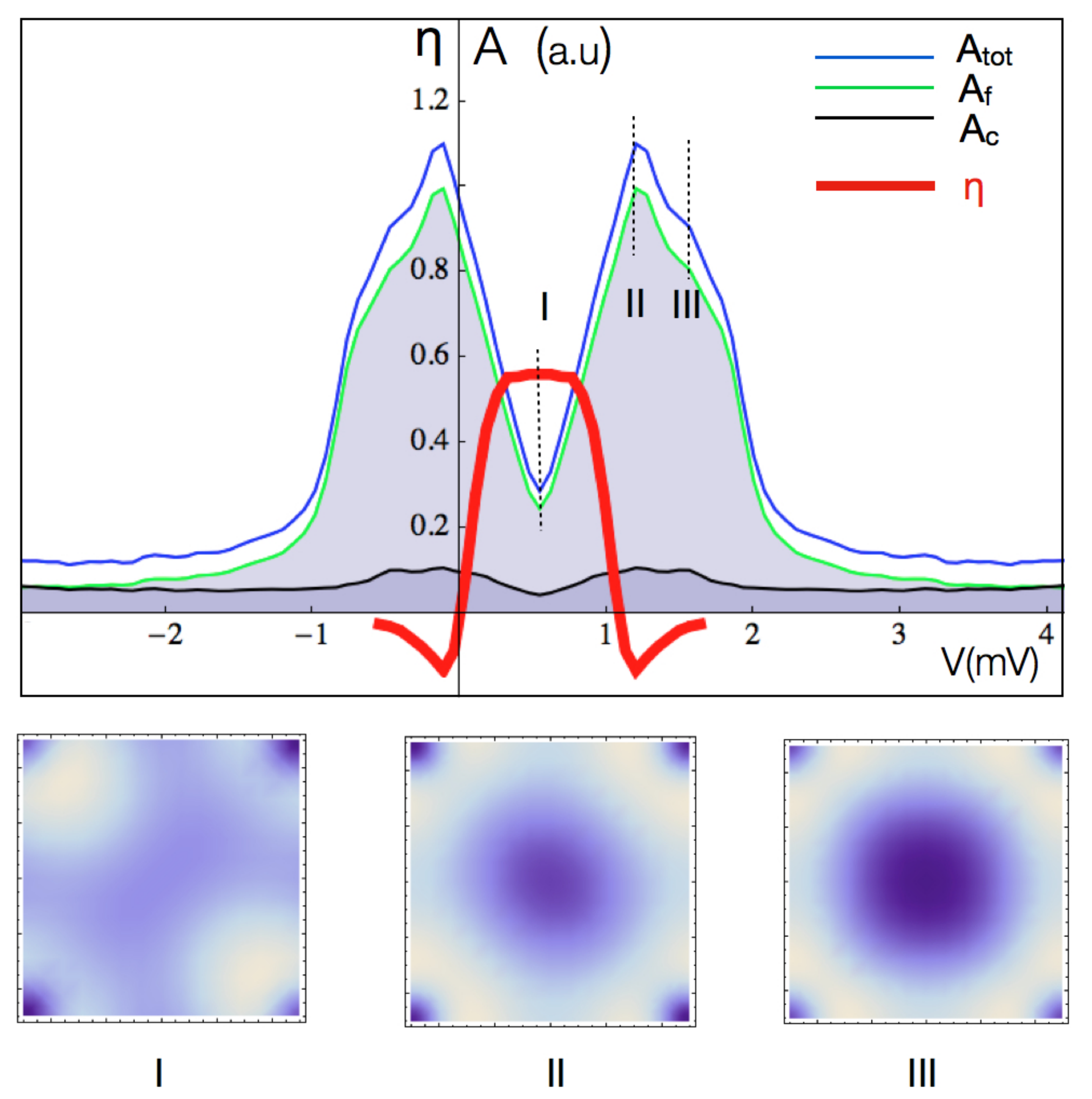}}
{\vspace{3cm}
\Large \sffamily Figure 4.}
\end{figure}
\newpage
\end{document}


\centerline{\Large\bf   
Supplementary Information}


\vskip 0.3in
\centerline{\bf Online material: Hastatic order in \urs
}
\vskip 0.4in
\centerline{Premala Chandra$^1$, Piers Coleman$^{1,2}$ and Rebecca Flint$^3$}
\centerline{\sl $^1$Center for Materials Theory,
Rutgers University, 136 Frelinghuysen Rd., Piscataway, NJ 08854-8019, USA}
\centerline{\sl $^2$Department of Physics, Royal Holloway, University
of London, Egham, Surrey TW20 0EX, UK.} 
\centerline{\sl $^3$Department of Physics, Massachusetts Institute for Technology, Cambridge, MA 02139-4307,USA}
\

\tableofcontents

\vfill \eject

\section{Absence of Ising symmetry in a Kramers doublet with
Tetragonal Symmetry.}

A central argument in the main paper is that the Ising symmetry of the
quasiparticles in \urs must derive from a non-Kramers doublet. This
section establishes that a Kramers doublet in a tetragonal
crystal lacks the selection rules required for an Ising symmetry 
without fine tuning.
A tetragonal crystal field Hamiltonian contains
terms of the form $J_\pm^4$ allowed by 
the four-fold symmetry in
the basal plane.
These terms cause a pure doublet $|\pm
M\rangle$(where $M$ is some $J_z \in \{-J,\ldots J\}$), to mix with 
states $\vert \pm M'\rangle $ differing by four units of angular momentum, 
where $M' = M -4 n$, and $n$ is any integer.  A magnetic
doublet in an tetragonal environment thus has the form
\begin{eqnarray}\label{l}
|\Gamma + \rangle & = & \sum_n a_n |M-4n\rangle\cr
|\Gamma - \rangle & = & \sum_n a_n |-M+4n\rangle. 
\end{eqnarray}
The coefficients $a_{n}$ may always be 
chosen to be real. Ising symmetry 
requires that the matrix elements 
\begin{equation}\label{sum}
\langle \Gamma + |J_\pm
|\Gamma -\rangle  = \sum_{n,n'}a_{n}a_{n'}\langle M - 4n \vert 
J_{\pm}\vert -M + 4n'\rangle
\end{equation}
vanish. 
In the absence of fine-tuning ($a_n = 0$ for all $n \neq 0$), 
this implies a selection rule 
$\langle M-4n|J_\pm
|-M +4n' \rangle =0$. 
Such terms vanish if $M-4n \neq  -M+4n'\pm 1$, or $M \neq  2
(n-n')\pm \frac{1}{2}$. Since $n$ and $n'$ are integers, $M$ must be
an integer. {For } any
half-integer $M$, corresponding to a Kramers doublet, {the
selection rule is absent and the ion}  will
develop a generic basal plane moment. 
{The fine-tuned case will produce an Ising Kramers
doublet, but corresponds to the complete absence of tetragonal mixing, highly unlikely
in a tetragonal environment. }
However, for integer $M$,
corresponding to a non-Kramers doublet, {the selection rule 
$\langle M-4n \pm |J_\pm
|\Gamma -M +4n' \rangle =0$ causes }every term in the above sum
(\ref{sum}) {to}
vanish {so} the transverse moment is necessarily zero, yielding
perfect  Ising symmetry. 

By contrast, in a hexagonal system like CeAl$_3$, the crystal field
Hamiltonian contains terms of the form $J_\pm^6$.  Such terms again
mix a pure doublet $|\pm M\rangle$ with terms $|\pm M'\rangle$, where
$M' = M - 6 n$, integer $n$.  However, for $J < 7/2$, there are no
choices of $M$ and $M'$ that differ by $6$, and thus there will be two
Ising doublets for the Ce, $J=5/2$ case: $\Gamma_8 = |\pm 5/2$ and
$\Gamma_9 = |\pm 3/2\rangle$.  Either of these Kramers doublets could
undergo a {\sl single channel} Ising Kondo effect\cite{Goremychkin00,
sikkema96}, which will differ substantially from the two-channel Kondo
physics associated with a non-Kramers doublet.

\section{Estimate of $T_K$ for ${\rm URu_2Si_2}$}

In our mean field theory, all Kondo behavior develops at the hidden
order transition.  Incorporating Gaussian fluctuations should suppress
the hidden order phase transition, $T_{HO}$, while allowing many of
the signatures of heavy fermion physics, including the quenching of
the spin entropy and the heavy mass to develop at a higher crossover
scale, $T_K$.  While the coherence temperature estimated from the
resitivity, $T^* \approx 70K$ is much larger than the hidden order
temperature, $T_{HO} = 17.5K$, an entropic estimate of the Kondo
temperature, $S(T_K) = \frac{1}{2} R \log 2$ gives an effective Kondo
temperature not much larger than $T_{HO}$.  There is considerable
uncertainty in the entropy associated with the development of hidden
order, $S(T_{HO})$, due to difficulties subtracting the phonon and
other non-electronic contributions, leading to estimates ranging from
$.15 R\log 2$\cite{Jaime02} to $.3 R \log 2$\cite{Palstra85}.  If we
take a conservative estimate of $S(T_{HO}) = .2 R\log 2$, and the
normal state $\gamma = 180 {\rm mJ/mol K^2}$\cite{Palstra85}, $S(T_K)
= .2 R \log 2 + \int_{T_{HO}}^{T_K} \gamma dT = \frac{1}{2} R \log 2$
yields $T_K = 27$K, much lower than the coherence temperature seen in
the resistivity.

\section{Landau theory for hastatic order.}\label{}
\figwidth=12cm 
\fgb{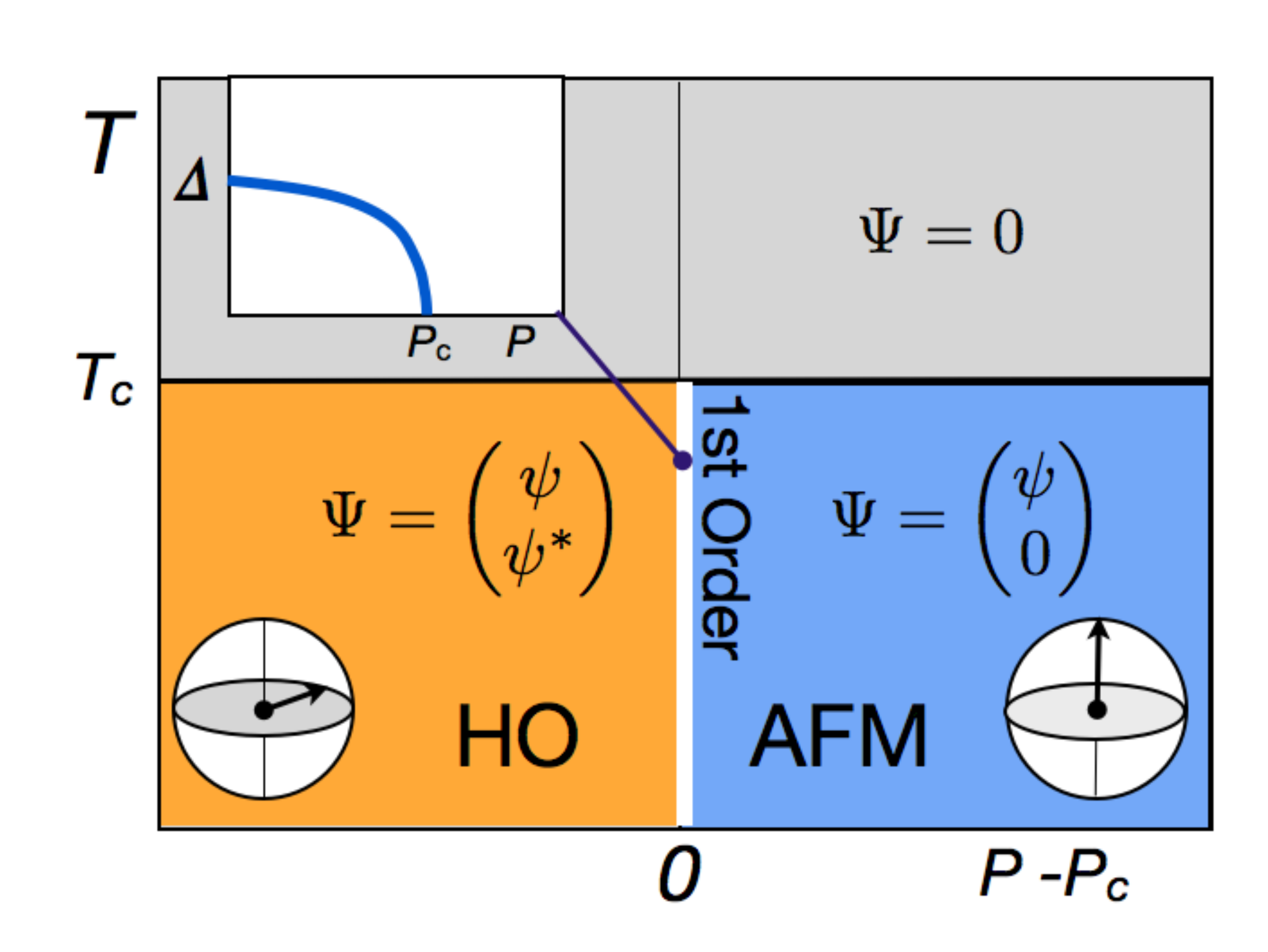}{landy}{Global phase diagram predicted by
Landau theory.}

\subsection{Landau theory in zero field}

The most general Landau functional for the free energy density of a
hastatic state with a spinorial order parameter $\Psi $ as a function
of pressure and temperature is 
\begin{equation}\label{}
f[\Psi] = 
\alpha (T_{c}-T)\vert \Psi \vert^{2}+ \beta \vert \Psi \vert^{4} - 
\gamma (\Psi \dg \sigma_{z}\Psi )^{2}
\end{equation}
where $\gamma = \delta(P-P_{c})$ is a pressure-tuned anisotropy term and 
\begin{equation}\label{}
\Psi  = r \left(\begin{matrix}
\cos (\theta/2 )e^{i\phi /2}\cr
\sin(\theta/2)e^{-i\phi /2}\end{matrix}
\right), 
\end{equation}
where $\theta $ is the disclination of $\Psi\dg  \vec{\sigma }\Psi $
from the c-axis. 
Using this expression for $\Psi $,
\begin{equation}\label{}
f= - \alpha  (T-T_{c})r^{2}+ \beta  r^{4}- \gamma r^{4 }\cos^2 \theta.
\end{equation}
If $P< P_{c}$, then $\gamma <0$ and the minimum of the free energy
occurs for $\theta  = \pi/2$, corresponding to the hidden order state
ordered state. By contrast, if $P>P_{c}$, then $\gamma > 0$ and the
minimum of the free energy occurs at $\theta  = 0, \pi$, corresponding
to the antiferromagnet.  The ``spin flop'' in $\theta $ at $P=P_{c}$
corresponds to a first order phase transition between the hidden order
and antiferromagnet (See Fig. \ref{landy})

To study the soft modes of the hastatic order, we need to generalize
the Landau theory to a time-dependent Landau Ginzburg theory for the
action, with action $S= \int L dt d^{3}x$, where the Lagrangian
\[
-L[\Psi ]  = f [\Psi] + \rho \left(|\nabla \Psi |^{2}  - c^{-2}|\dot{\Psi }|^{2}\right),
\]
and $\rho $ is the stiffness.
Expanding $\Psi$ around its equilibrium value $\Psi_{0}$, taking $\phi
=0$ for convenience and writing
\begin{equation}\label{}
\Psi (x,t)=  \Psi_{0}e^{i\delta \theta (x)\sigma_{y}/2} = 
(1 + i/2\sum_{q}\delta \theta (q)e^{i \vec{ q}\cdot
\vec{x}-\omega t}\sigma_{y})\Psi_{0} .
\end{equation}
This gives rise to a change in 
$\Psi \dg \sigma_{z}\Psi = \hat x |\Psi_{0}|^{2}+ \delta \theta (x)
\hat z |\Psi_{0}|^{2}$
corresponding to a fluctuation in the longitudinal magnetization.
This rotation in $\Psi $ does not affect the first two isotropic 
terms in $f[\Psi ]$. 
The variation in the action is then given by 
\begin{equation}\label{}
\delta S = \rho |\Psi_{0}|^{2} \sum_{q} \vert  \theta (q)\vert^{2}
\left(\vec{q}^{\ 2}- \frac{\omega^{2}}{c^{2}}+\frac{2\delta}{\rho}  (P_{c}-P)|\Psi_{0}|^{2} \right)
\end{equation}
The dispersion is therefore 
\begin{equation}\label{}
\omega^{2} = (c q)^{2}+ \Delta^{2}
\end{equation}
where 
\begin{equation}\label{}
\Delta^{2}= \frac{2 \delta (P_{c}-P)}{\rho }|\Psi_{0}|^{2},
\end{equation}
so that even though the phase transition at $P=P_{c}$ is first order, 
the gap for longitudinal spin fluctuations is
\[
\Delta \propto  |\Psi_{0} | \sqrt{P_{c}-P}.
\]
Since $dP_c/dT_c$ is finite, close to the transition, $\sqrt{P_c-P} \approx \sqrt{dP_c/dT_c(T-T_c)}$, and $\Delta \propto \sqrt{T-T_c}$.  Inelastic neutron scattering experiments can measure this gap a function of temperature at a fixed pressure where there is a finite temperature first order transition.  The iron-doped compound, URu$_{2-x}$Fe$_x$Si$_2$ can provide an attractive alternative to hydrostatic pressure, as iron doping acts as uniform chemical pressure and tunes the hidden order state into the antiferromagnet\cite{butch}.

\subsection{Landau Theory in Magnetic Field: Nonlinear Susceptibility}

The origin of the large c-axis nonlinear susceptibility anomaly in \urs\cite{ramirez94} has been a long-standing mystery.  It has been understood phenomenologically within a Landau theory as a consequence of a large $\Psi^2 B_z^2$ coupling of unknown origin\cite{ramirez94,newrefschi3}.  Here we show that the Landau theory of hastatic order will contain just such a term.

While the conduction electrons couple isotropically to an applied field, the non-Kramers doublet linearly couples only to the z-component of the magnetic field, $B_{z}= B\cos \theta$, which splits the doublet as it begins to suppress the Kondo effect.  When we include the effect of the magnetic field in the Landau theory, we obtain:
\begin{equation}\label{}
f[\Psi] = \bigl [\alpha (T_{c}-T)- \eta_z B_{z}^{2} -\eta_\perp B_\perp^2\bigr ]\Psi^{2} + \beta
\vert \Psi |^{4}  +\gamma (\Psi \dg \sigma_{z}\Psi )^{2},
\end{equation}
where we shall show below that the coefficient of the $\Psi^2 B_z^2$ term, $\eta_z$ goes as $\rho/T^2_{HO}$, where $\rho$ is the conduction electron density of states, while the coefficient of the $\Psi^2 B_\perp^2$ term, $\eta_\perp$ is of order $\rho/D^2$, where $D$ is the conduction electron bandwidth.  Minimizing this functional with respect to $\Psi$, we obtain
\begin{equation}\label{}
f = - \frac{1}{4 \beta }\left[\alpha (T_{c}-T)- \eta_z (B\cos \theta)^{2} - \eta_\perp (B \sin \theta)^2 \right]^{2}.
\end{equation}
Following the arguments of \cite{newrefschi3}, 
we can calculate the jump in the specific heat 
$\Delta C_{v}$ and the susceptibility
and non-linear susceptibility anomalies $d\chi_{1}/dT$ and $\Delta\chi_{3}$
respectively, to find
\begin{eqnarray}\label{l}
\frac{\Delta C_{V}}{T_{HO}} &=&
\frac{\alpha^{2}}{2 \beta }\\
\frac{d\chi_{1}}{dT} &=& - \frac{\alpha }{2 \beta }\left(\eta_z \cos^2 \theta + \eta_\perp \sin^2 \theta\right) \approx - \frac{\alpha \eta_z}{2 \beta } \cos^2 \theta \\
\Delta \chi_{3} &= & 
 \frac{6}{\beta } \left(\eta_z \cos^2 \theta + \eta_\perp \sin^2 \theta\right)^2 \approx \frac{6\eta_z^2}{\beta }\cos^4\theta
\end{eqnarray}
where $\chi_{1}$ and $\chi_{3}$ are the anomalous
components of the susceptibility that develop at $T_{HO}$.  These results
show that $\chi_{1}$ and $\chi_{3}$ will exhibit a giant Ising
anisotropy.  The thermodynamic relation
\begin{equation}\label{}
\frac{\Delta  C}{T} \chi_{3}= 12 \left(\frac{ d \chi_{1}}{dT} \right)^{2}
\end{equation}
 still holds. 

\subsubsection{Estimating $\eta_z$ and $\eta_\perp$}

To illustrate this simple Landau theory, we will calculate
$\vec{\eta}$ in a simplified model: we will neglect the momentum
dependence of both the f-level and the hybridization and take the
hastatic order to be uniform.  None of these assumptions qualitatively
changes the results.  The $|\Psi|^2$ coefficient is calculated from
the microscopic theory (see the next section) by expanding the action,
$S = -\tr \log\left[ 1 - \mathcal{F}_0 (V\Psi) \mathcal{G}_0 (V
\Psi\dg)\right]$ in $\Psi$, where $\mathcal{F}_0 = (i\omega_n -
\lambda - g_f \mu_f B_z \sigma_3)^{-1}$ and $\mathcal{G}_0 =
(i\omega_n - \epsilon_\bk-g/2 \vec{B}\cdot \vec{\sigma})^{-1}$ are the
bare $\chi$ and conduction electron Green's functions (remember,
$\chi$ are the fermions representing the non-Kramers doublet).  $V$
represents the hybridization matrix elements, which are
momentum-independent here, and proportional to the unit matrix.  Note
that while the conduction electrons are isotropic, the $\chi$'s are
perfectly Ising.
The coefficient of $|\Psi|^2$ is then,
\upit=-0.4in
\bea
\raiser{\frm{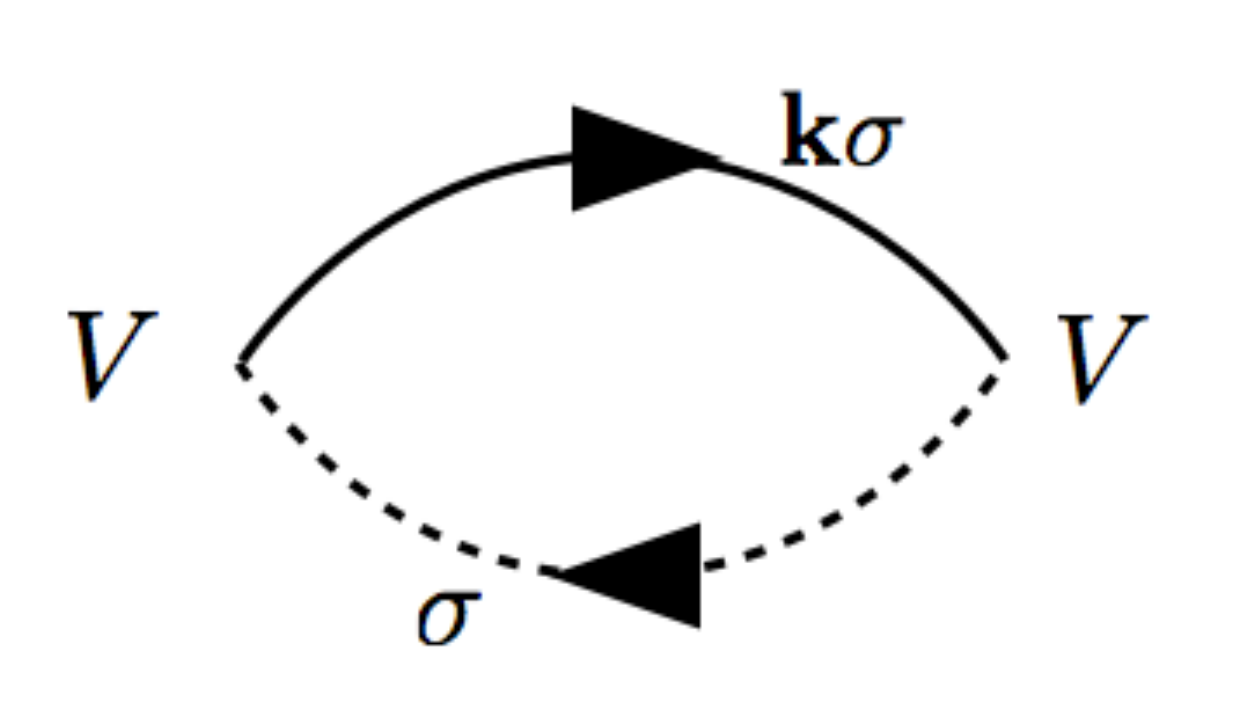}}=
V^2 T \sum_{i \omega_n} \sum_{\bk\sigma} \frac{1}{i\omega_n - \epsilon_{\bk \sigma}}\frac{1}{i\omega_n - \lambda_\sigma},
\eea
where $\epsilon_{\bk \sigma} = \epsilon_\bk - g/2\vec{\sigma}\cdot
\vec{B}$ and $\lambda_\sigma = \lambda - g_f \mu_f \sigma B_z$ are the
dispersions in field. Performing the Matsubara sum, we obtain
\bea
V^2 \sum_{\sigma}\int_{-\infty}^{\infty}d\epsilon \mathcal{D}(\epsilon) \frac{\tanh \frac{\epsilon_{\bk \sigma}}{2T} - \tanh \frac{\lambda_\sigma}{2T}}{2 (\lambda_\sigma - \epsilon_{\bk \sigma})} = \rho V^2 \sum_{\sigma}\int_{-D}^{D}d\epsilon \frac{\tanh \frac{\epsilon-g/2\vec{\sigma}\cdot \vec{B}}{2T} - \tanh \frac{\lambda_\sigma}{2T}}{2 (\lambda_\sigma - \epsilon + g/2 \vec{\sigma}\cdot \vec{B})},
\eea
where we approximated the conduction electron density of states as a constant, $\rho$ within the bandwidth, $2D$.
Let us first calculate $\eta_\perp$, quantizing the field along the x-direction and taking g = 2,
\bea
\eta_\perp 
=
\raiser{\frm{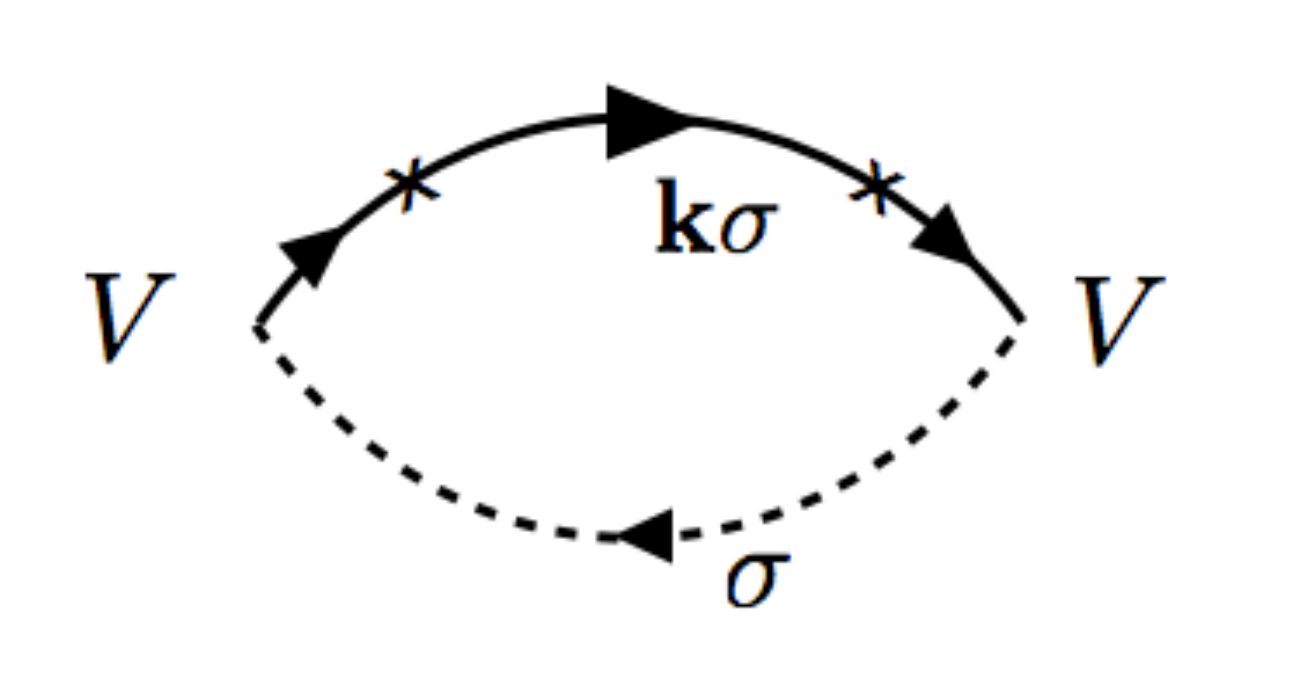}}
= -\rho V^2 \sum_\sigma \int_{-D}^{D}d\epsilon \left.\frac{\partial^2}{\partial B_\perp^2}\frac{\tanh \frac{\epsilon-\sigma B_\perp}{2T} - \tanh \frac{\lambda}{2T}}{2 (\lambda- \epsilon + \sigma B_\perp)}\right|_{B_\perp = 0}.
\eea
As the integrand is a function of $\epsilon - \sigma B$, the integral is straightforward.  And as $D \gg \lambda, T$, the dominant term will be:
\bea
\eta_\perp = -\rho V^2 \left.\left( \frac{{\rm sech}^2\frac{\epsilon}{2T}}{4 T(\epsilon-\lambda)}+\frac{\tanh \frac{\epsilon}{2T}-\tanh\frac{\lambda}{2T}}{(\epsilon-\lambda)^2}\right)\right|_{-D}^{D} = \frac{\rho V^2}{D^2}.
\eea
$\eta_z$ will have three contributing terms: one purely from the conduction electrons that is $\eta_\perp$, one arising from cross terms between the conduction and f-electrons, and finally one solely from the f-electrons that dominates the other two.  We shall focus on this last term,
\bea
\eta_z =
\raiser{\frm{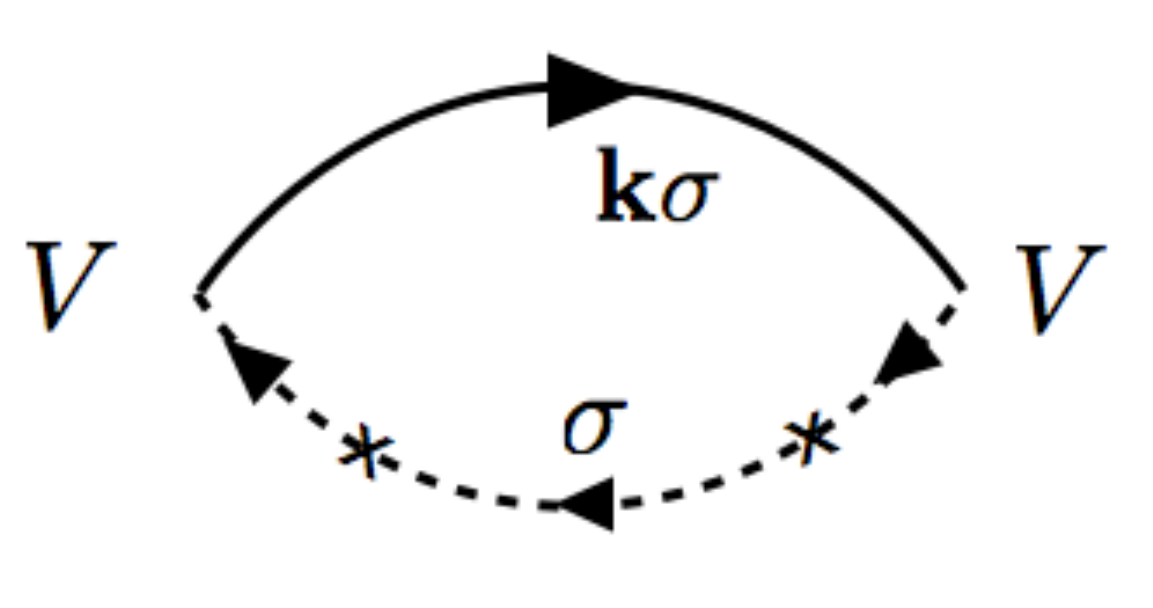}}=
-\rho V^2 \sum_\sigma \int_{-D}^{D} d\epsilon \left.\frac{\partial^2}{\partial B_z^2}\frac{\tanh \frac{\epsilon}{2T} - \tanh \frac{\lambda-g_f \mu_f \sigma B_z}{2T}}{2 (\lambda- \epsilon-g_f \mu_f \sigma B_z)}\right|_{B_z = 0}.
\eea
This integral cannot be done analytically at finite temperature, so we take $T\rarrow 0$.
\bea
\eta_z \approx -\rho V^2\left.\frac{\partial^2}{\partial B_z^2} \sum_\sigma \int_{-D}^0 \frac{1}{2(\lambda_\sigma - \epsilon)} \right|_{B_z = 0} = \frac{\rho V^2}{\lambda^2} - \frac{\rho V^2}{(D+\lambda)^2} = \frac{\rho V^2}{T_{HO}^2},
\eea
as $\lambda = T_{HO}$ at zero temperature.  So $\eta_\perp/\eta_z = \frac{T_{HO}^2}{D^2}$.  
Using a conservative of
$T_{H0}/D\sim 1/30$, we predict an anisotropy of 
about $900$ in $d\chi_{1}/dT$ and nearly $10^6$ in
$\Delta \chi_{3}$.  However, in a realistic model, there will be f-electron contributions to $\eta_\perp$ involving fluctuations to excited crystal field states that may reduce the anisotropy somewhat.
The important point here is that the anisotropies will
be orders of magnitude larger than the single ion anisotropy in
$\chi_{1}$ (approximately $3$), 
and furthermore, that they will develop exclusively at the hidden order
transition.  

\section{Two channel Anderson lattice model for \urs}

Our model describes a lattice of $U$ ions immersed in a conduction sea
of electrons. 
We take the low energy configuration of each $U$ ion 
to be a $5f^2$ $\Gamma_5$ non-Kramers doublet. 
All energies are measured relative
to the energy of the isolated doublet. 
For
simplicity, in our model we take the dominant valence fluctuation
channel to be $5f^{2}\rightleftharpoons 5f^{1}$.  Particle-hole
symmetry can be used to formulate the equivalent model with
fluctuations into a $5f^{3}$ Kramers doublet. 
The full model is then written 
\begin{equation}\label{}
H = \sum_{ \bk \sigma }\epsilon_{\bk }c\dg_{\bk \sigma }c_{\bk \sigma}
+ \sum_{j}\left(H_{VF} (j)+ H_{a} (j) \right)
\end{equation}
where $c\dg_{\bk \sigma }$ creates a conduction electron of 
momentum
$\bk $ spin $\sigma $, with energy $\epsilon_{\bk }$, 
\begin{equation}
\label{VF} H_{VF}(j) = V_6 \psi_{\Gamma_6 \pm}\dg(j) |\Gamma_7^+
\pm\rangle \langle \Gamma_5 \pm| + V_7 \psi_{\Gamma_7 \mp}\dg(j)
|\Gamma_7^+ \mp \rangle\langle \Gamma_5 \pm| + \mathrm{H.c.}
\end{equation}
describes the valence fluctuations between the $\Gamma_{5}$ doublet
and an excited $\Gamma_{7}$ Kramers doublet while
\begin{equation}\label{}
H_a(j) = \Delta E \sum_\pm|\Gamma_7 \pm\rangle \langle \Gamma_7 \pm| 
\end{equation}
is the atomic Hamiltonian. 

In our model, we choose the $\Gamma_{7}$ to
be the lowest excited state; in this case the 
the valence fluctuations are determined by the decomposition,
\begin{equation}
|\Gamma_7 \pm\rangle = \alpha\psi\dg_{6 \mp}|\Gamma_5 \pm\rangle + \beta\psi\dg_{7 \mp}|\Gamma_5 \mp \rangle,
\end{equation}
where $\psi\dg_{\Gamma \alpha} = \sum_\bk c\dg_{\bk\beta} \left[\Phi_{\Gamma
\bk}\right]_{\alpha \beta} \mathrm{e}^{-i \bk\cdot \bR_j}$ creates
a conduction electron in a Wannier state with $j=5/2$ and symmetry
$\Gamma$ ($=\Gamma_6,\Gamma_7$) localized around the uranium atom at
site $j$.

For a plane wave, the form factors are given by 
$\left[\Phi_{\Gamma \bk}\right]_{\alpha \beta} = y^{\Gamma}_{\alpha
\beta} 
(\bk )$, where
\[
y^{\Gamma}_{\alpha \beta}(\bk )
= Y_{3m-\frac{\alpha }{2}} (\hat
\bk ) 
\langle 3m-\frac{\alpha }{2}, \frac{1}{2} \frac{\alpha }{2}\vert 5/2 m \rangle 
\overbrace {\langle m\vert \beta\rangle }^{a_{m}}
\]
In \ursp,  the uranium 
atoms are located on a body centered tetragonal lattice (bct) at
relative locations, $\bR_{NN} = ( \pm a/2, \pm a/2, \pm c/2)$, while the f-electrons hybridize with conduction electrons located around the silicon atoms, ${\bf{a}}_{NN} = ( \pm a/2, \pm a/2, \pm z)$ where $z = .371 c$ is the height of the silicon atom above the $U$ atom\cite{Oppeneer10}.
The form-factor is then, 
\begin{equation}\label{}
\left[\Phi_{\Gamma \bk}\right]_{\alpha \beta} = \sum_{\{{\bf{R}}_{NN}, {\bf{a}}_{NN}\}
} e^{-i \bk
\cdot{\bf R}_{NN}} y^{\Gamma}_{\alpha \beta} 
({\bf{a}}_{NN})
\end{equation}
Notice that this function has the following properties:
$\left[\Phi_{\Gamma \bk+{\bf G}}\right]_{\alpha \beta} 
=\left[\Phi_{\Gamma \bk}\right]_{\alpha \beta} $ and 
$\left[\Phi_{\Gamma \bk+\bQ}\right]_{\alpha \beta} 
=-\left[\Phi_{\Gamma \bk}\right]_{\alpha \beta}$. 

To cast the Hamiltonian as a field theory, we introduce a slave
boson/slave fermion representation, $|\Gamma_7 \sigma\rangle\langle
\Gamma_5 \alpha| = b\dg_\sigma \chi _\alpha$, where
$b\dg_\sigma|\Omega\rangle$ represents the occupation of $5f^1$, and
$b_\sigma$ carries a positive charge and  $\chi_\alpha\dg |\Omega\rangle$
represents the $5f^2$ state.  
The valence fluctuation term in the Hamiltonian then takes the form 
 \bea \label{VF3} H_{VF1}(j)  =  \sum_\bk
c\dg_{\bk\sigma} \hat{\mathcal{V}}^{(1)}_{\sigma \alpha}(\bk,j) \chi
_\alpha(j) \mathrm{e}^{-i\bk\cdot \bR_j} + \mathrm{H.c.}
\eea
Introducing $\hat B\dg = \zmatrix{\hat\Psi\dg_\up}{0}{0}{\hat\Psi\dg_\dw}$ (where we use $\hat\Psi\dg_\sigma$ instead of the traditional $b\dg_\sigma$ to reinforce that these are indeed order parameters), we can write the hybridization compactly,
\bea
\hat{\mathcal{V}}^{(1)}(\bk,j)  = V_6 \Phi_{\Gamma_6} B\dg_j +   V_7
\Phi_{\Gamma_7^-} \hat B\dg_j\sigma_1.
\eea

In the mean field theory, we replace $\hat {\cal V} (\bk,j)
\rightarrow \langle \hat {\cal V} (\bk,j)\rangle $, 
replacing $\langle \hat \Psi\dg_{\sigma}\rangle  = \Psi^{*}_{\sigma }$.
We consider the configuration
\begin{eqnarray}\label{l}
\langle \hat B\dg_{j}\rangle  &=& |\Psi| \left(\begin{matrix}
e^{i (\bQ \cdot \bR_{j}+\phi) /2}
& \cr
&e^{-i (\bQ \cdot \bR_{j}+ \phi) /2}
\end{matrix} \right) = |\Psi| U_{j}
\end{eqnarray}
where $U_{j}$ is a unitary matrix. and $\bQ= (0,0,\frac{2
\pi}{c})$ is the common wavevector for hastatic order, chosen
to match the antiferromagnet, and $\phi  = \pi/ 4$ is chosen 
to match the susceptibility anisotropy.  
It is convenient to redefine $\tilde{\chi}_{j}= U_{j} \chi_{j}$.
With this device, the spatial dependence of $\langle \hat B_{j}\dg\rangle$ 
is absorbed  into the redefined f-electrons, so that 
$\hat B\dg_{j}\chi_{j} = |\Psi|\tilde{\chi}_{j}$
and 
\begin{equation}
\hat B\dg_{j}\sigma_1 \chi_{j} = |\Psi| (U_j\sigma_{1}U\dg_{j})\tilde{\chi}_{j} = |\Psi| (\hat {\bf{n}}\cdot \vec{\sigma})
e^{i \bQ \cdot \bR_{j}}\tilde{\chi}_j, \quad  \hat {\bf{n}} = \frac{1}{\sqrt{2}}(\hat {\bf{x}}+
\hat {\bf{y}}
).
\end{equation}
In this gauge, the $\Gamma_6$ hybridization is uniform while the
$\Gamma_{7}^{-}$ hybridization is staggered.  
We
write ${\cal V}_{6} (\bk ) = |\Psi|V_{6} \Phi_{\Gamma_6}$, and ${\cal
V}_{7} (\bk ) =|\Psi| V_{7}\Phi_{\Gamma_7}(\hat {\bf{n}}\cdot
\vec{\sigma})$.

In the slave formulation, the atomic Hamiltonian is $H_a(j) = \Delta E
\sum_\sigma \Psi\dg_{j\sigma} \Psi_{j\sigma}$.  The introduction of slave
bosons and fermions to represent the Hubbard operators requires a
constraint to maintain one particle per site, $\lambda_j \left(
\sum_\sigma \Psi\dg_{j\sigma} \Psi_{j\sigma} + \sum_\alpha\chi\dg_{j\alpha}
\chi_{j\alpha} - 1\right)$. 

We take a simplified model of the
conduction electron hopping, treating them as s-wave electrons located
at the $U$ site, hopping on a bct lattice with dispersion
\begin{equation}
\epsilon_\bk = -8 t \cos \frac{k_x a}{2}\cos \frac{k_y a}{2}\cos \frac{k_z c}{2} - \mu.
\end{equation} 
We do, however, want to capture the essential characteristics of the
\urs bandstructure - namely nesting between an electron Fermi surface
about the zone center and a hole Fermi surface at $\bQ$\cite{Oppeneer10}.  In order to
favor a staggered hybridization, and to match up with ARPES
experiments suggesting a heavy f-band\cite{Santander09}, we take the hole Fermi surface
to be generated from a weakly dispersion $\chi$ band.  This f-electron
hopping will be naturally generated by hybridization fluctuations
above $T_{HO}$, effectively where $\langle \hat B\dg \hat B\rangle \neq 0$ while
$\langle \hat B \rangle = 0$.  A large $N$ expansion of this problem would capture these
fluctuation effects, but is overly complicated for this problem so we
put this dispersion in by hand, $\epsilon_{f\bk} = -8 t_f \cos \frac{k_x
a}{2}\cos \frac{k_y a}{2}\cos \frac{k_z c}{2}$.  

So to summarize, our mean-field Hamiltonian is,
\begin{equation}
H = \sum_\bk \epsilon_\bk c\dg_{\bk \sigma} c_{\bk \sigma} + \sum_\bk tf_{\bk} \chi\dg_{\bk \eta} \chi_{\bk \eta} + \sum_j (\Delta E +\lambda_j) \Psi\dg_{j\sigma} \Psi_{j\sigma} + \lambda_j \left(\chi\dg_{j\eta} \chi_{j\eta} - 1\right) + \sum_j H_{VF}(j).
\end{equation}
We rewrite this Hamiltonian  in matrix form
\begin{eqnarray}\label{thematrix}
H & = & \sum_\bk \left( c\dg_\bk, c\dg_{\bk+\bQ}, \chi\dg_\bk,
\chi\dg_{\bk+\bQ} \right) 
\overbrace {\left(\begin{matrix}\epsilon_\bk & 0 & \mathcal{V}_6(\bk) & \mathcal{V}_7(\bk)\cr 0 & \epsilon_{\bk+\bQ} & -\mathcal{V}_7(\bk) & -\mathcal{V}_6(\bk)\cr
\mathcal{V}_6\dg(\bk) & -\mathcal{V}_7\dg(\bk) & \lambda_\bk & 0\cr
\mathcal{V}_7\dg(\bk) & -\mathcal{V}_6\dg(\bk) & 0 &
\lambda_{\bk+\bQ}\end{matrix}\right)
}^{{\cal H}_{\alpha \beta } (\bk )}
\left(\begin{array}{c}
c_\bk \\ c_{\bk+\bQ} \\ \chi_{\bk} \\ \chi_{\bk+\bQ}
\end{array} \right)\cr
&&  + \sum_j \left[(\Delta E + \lambda) \Psi\dg_{j\sigma} \Psi_{j\sigma} - \lambda\right].
\end{eqnarray}
where we have suppressed spin indices, made the assumption that
$\lambda_j = \lambda$ is uniform, equivalent to enforcing the
constraint on average, introduced $\lambda_\bk = \lambda -\epsilon_{f\bk}$, and
used the simplification that $\bQ$ is half a reciprocal lattice
vector, making $\mathcal{V}(\bk+\bQ) = -\mathcal{V}(\bk)$, as shown
above.

In the absence of particle-hole symmetry, this Hamiltonian cannot be diagonalized analytically, and must be done numerically, giving a set of four doubly degenerate bands, $E_{\bk \eta}$.  The mean field free energy is,
\begin{equation}
F[b,\lambda] = -\frac{\beta^{-1}}{2}\sum_{\bk, \eta}\log\left[1 +\mathrm{e}^{-\beta E_{\bk\eta}}\right] + \mathcal{N}_s \left[2(\Delta E + \lambda) |\Psi|^2 - \lambda\right],
\end{equation}
where $\beta = (k_B T)^{-1}$.  The mean field parameters, $|\Psi|$ and
$\lambda$ are obtained by numerically finding a stationary point of
the free energy that minimizes $F$ with respect to $|\Psi|$ and maximizes
it with respect to $\lambda$.  A plot of $|\Psi|$ and $\lambda$ as a
function of temperature is shown in Figure 1 C, for the parameters
used to calculate $\chi_{xy}(T)$ and $m(T)$ for Figure in the main
paper.  The parameters are as follows: t = 12.5 meV is taken to match the magnitude of $\chi_{xy}$ from the torque magnetometry data\cite{Okazaki11}.
$\mu/t = -.075$ gives the slight particle-hole asymmetry essential to reproduce the flattening of $\chi_{xy}$
at low temperatures, and has also been adjusted so that $\mu+\lambda = 0$ at $T=0$ for consistency with the dI/dV calculations; 
$t_f/t = -.025$ gives a weak f-electron
dispersion; the crystal field angle $\xi = .05$ is taken to be small,
as it is in CeRu$_2$Si$_2$\cite{Haen92}; $V_6/V_7 = 1$ is arbitrary; and finally
$V^2/\Delta E = 2 t$ is chosen to give $2|\Psi|^2 \approx 15\%$ mixed
valency.  The shape of the susceptibility curve is quite sensitive to
the particle-hole asymmetry and degree of mixed valency, but not to
the other parameters.  The internal angle, $\phi$ controls the
susceptibility anisotropy and the orientation of the magnetic moments.
$\phi = \pi/4$ reproduces $\chi_{xy} \neq 0$, $\chi_{xx} = \chi_{yy}$.

\figwidth=12cm 
\fgb{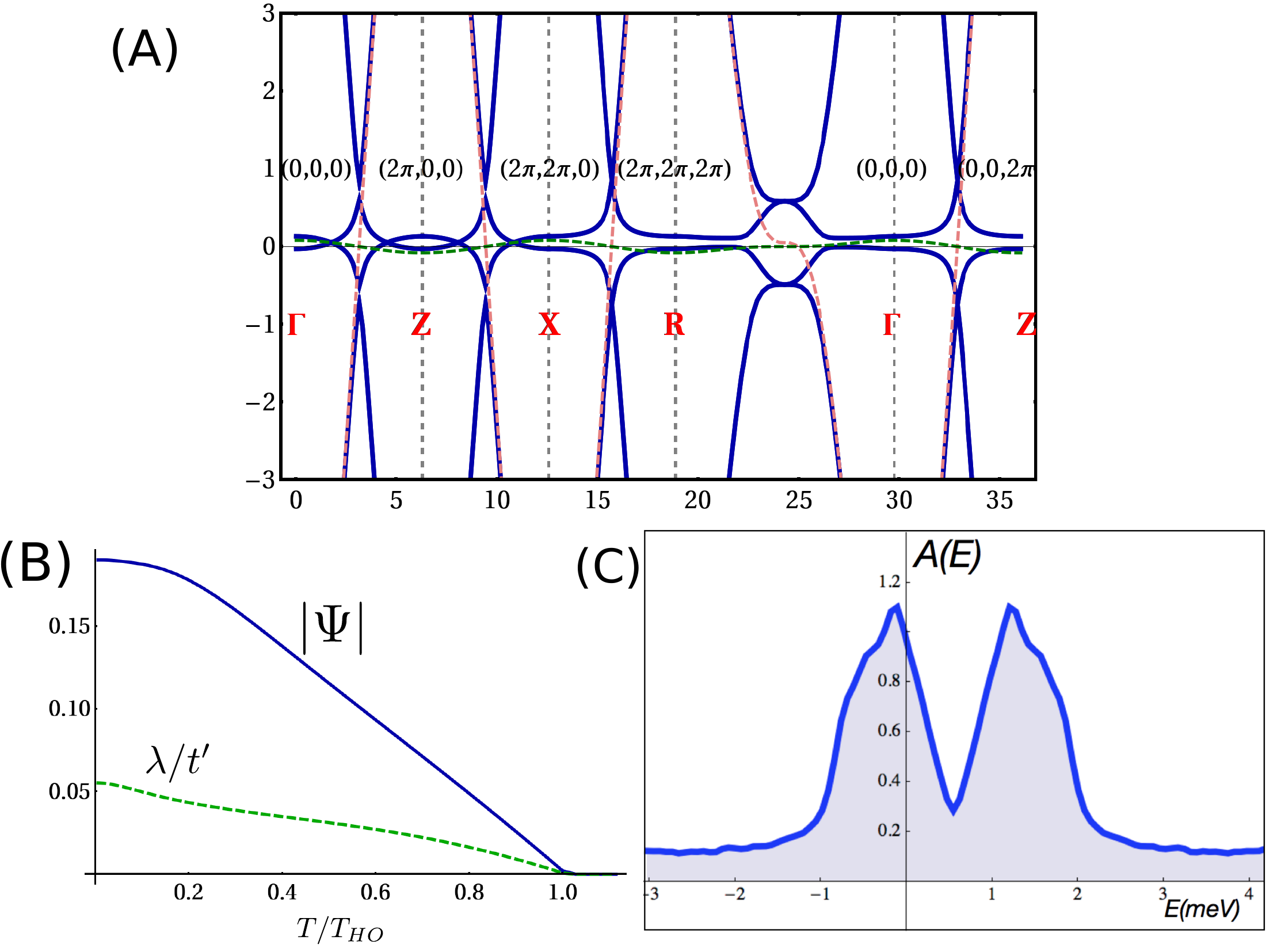}{etc}{(A) Band structure of the
hastatic order is shown in solid blue, while the bare conduction (red)
and f (green) bands are dashed. (B) Mean field parameters $|\Psi|$ and
$\lambda$ as a function of temperature. (C) Density of states in
hastatic order, for the region close to the Fermi energy containing
the hybridization gap.}

It is important to note that the hastatic state is promoted by a
partial nesting of the normal state Fermi surface along $\bQ = (100)$,
leading to the partial gapping of the Fermi surface seen in the
density of states (Fig. 2C).  However, even for perfect nesting, there
will still be gapless excitations, as the hybridization gap,
$\mathcal{V}_6\dg \mathcal{V}_7$ contains a node along the c-axis (due
to a c-axis node in $\mathcal{V}_7$).  In this sense (and this sense
only), the hastatic state is similar to the Ikeda-Miyake model\cite{ikeda96}
of the failed Kondo insulators CeNiSn and CeRhSb, which also
predicts a gapless state.

Above $T_{HO}$ in \ursp, there are magnetic fluctuations seen at two
momentum transfers: at the commensurate $\bQ = (1,0,0)$, which goes
soft at $T_{HO}$ and at the incommensurate $\bQ^* = (1\pm.4,0,0)$,
which remains at finite energy above both the HO and AFM
phases\cite{villaume}.  While we have chosen a simplified
bandstructure that captures the partial nesting at the commensurate
$\bQ$, it is likely that a more realistic bandstructure will contain
other energetically competitive nesting vectors that never develop due
to the stronger nesting at $\bQ$.  Fluctuations induced by these
nearby phases could be captured within an RPA treatment of a model
with a more realistic bandstructure.

For simplicity we have discussed the two channel Anderson model involving fluctuations from a $5f^2$ $\Gamma_5$ ground state to $5f^1$ ($J=5/2$).
However, the more realistic case involves fluctuations to $5f^3$, whose low energy states have $J=9/2$, and are split into five Kramers doublets by the tetragonal crystal field,
\bea
|\Gamma_7^1 \pm\rangle & = & a |\pm 5/2\rangle + b|\mp 3/2\rangle \cr
|\Gamma_7^2 \pm\rangle & = & -b|\pm 5/2\rangle +a|\mp 3/2\rangle \cr
|\Gamma_6^{1,2,3} \pm \rangle & = & c^{1,2,3}|\pm 9/2\rangle + d^{1,2,3}|\pm 1/2\rangle+ e^{1,2,3}|\mp 7/2\rangle.
\eea
There are two generic situations: either a $\Gamma_7$ doublet will be lowest in energy, and the valence fluctuations will be determined by the overlap,
\begin{equation}
|\Gamma_7 \pm\rangle = \alpha\psi\dg_{6 \mp}|\Gamma_5 \pm\rangle + \beta\psi\dg_{7 \mp}|\Gamma_5 \mp \rangle,
\end{equation}
or a $\Gamma_6$ doublet will be lowest in energy, with the relevant overlap,
\begin{equation}
|\Gamma_6 \pm\rangle = \alpha \psi\dg_{7 \mp}|\Gamma_5 \pm\rangle + \beta\psi\dg_{6 \mp}|\Gamma_5 \mp \rangle,
\end{equation}
where the form-factors are as above.  In both cases fluctuations will involve conduction electrons in both $\Gamma_6$ and $\Gamma_7$ symmetries.  When the lowest excited state is a $\Gamma_7$, the valence fluctuation Hamiltonian is given by,
\begin{equation}
H_{VF3}(j) =  V_6 \psi_{j\Gamma_6 \mp}\dg |\Gamma_5 \pm\rangle \langle \Gamma_7 \pm| + V_7 \psi_{j\Gamma_7 \mp}\dg |\Gamma_5 \mp \rangle\langle \Gamma_7 \pm| + \mathrm{H.c.}.
\end{equation}
Now we can follow the same slave boson procedure as discussed previous,  where $\Psi\dg_\sigma$ now represents a $5f^3$ state and carries a negative charge, and the valence fluctuation term is the particle-hole conjugate of the $5f^1$ case, $c\dg_{\bk\sigma} \hat{\mathcal{V}}_{\sigma \alpha}(\bk,j) \chi^* _\alpha(j) \mathrm{e}^{-i\bk\cdot \bR_j}$.  If we perform a particle-hole transformation on $\chi$, we regain the Hamiltonian, (\ref{thematrix}) with only a sign change for $\lambda_\bk$.

\section{Magnetization and susceptibility calculation}

The uniform basal plane conduction electron magnetic susceptibility is
\bea
\chi^{xy} = -(g \mu_B)^2 T \sum_{i\omega_n} \sum_{\bk}\tr\left[\sigma^{x}\mathcal{G}^c(\bk,\bk+\bQ,i\omega_n)\sigma^{y}\mathcal{G}^c(\bk+\bQ,\bk,i\omega_n)\right],
\eea
while the staggered conduction electron magnetization is
\bea
\vec{m}_c(\bQ) = -g \mu_B T\sum_{i\omega_n} \sum_\bk \tr \vec{\sigma} \mathcal{G}^c(\bk,\bk+\bQ,i\omega_n)
\eea
In order to calculate these, we require
the full conduction electron Green's function, which can be found from the Hamiltonian
by integrating out the f-electrons,
\bea
\left[\mathcal{G}^c(\bk,i\omega)\right]^{-1} = \zmatrix{i\omega_n - \epsilon_\bk}{0}{0}{i\omega_n - \epsilon_{\bk+\bQ}} - \mathcal{V}_{\bk} \zmatrix{i\omega_n - \lambda_\bk}{0}{0}{i\omega_n - \lambda_{\bk+\bQ}}^{-1}\mathcal{V}\dg_\bk,
\eea
where 
\bea
\mathcal{V}_\bk = \zmatrix{\mathcal{V}_{6\bk}}{\mathcal{V}_{7\bk}}{-\mathcal{V}_{7\bk}}{-\mathcal{V}_{6\bk}}.
\eea

Using isospin, $\vec{\tau}$ to represent $\bk, \bk+\bQ$ space, we 
split the conduction electron energy, $\epsilon_\bk$ into
$\epsilon_{0\bk} = \frac{1}{2}\left(\epsilon_{\bk} +
\epsilon_{\bk+\bQ}\right)$, $\epsilon_{1\bk} =
\frac{1}{2}\left(\epsilon_{\bk} - \epsilon_{\bk+\bQ}\right)$ into the
particle-hole symmetric and antisymmetric parts (and similarly with
$\lambda_{0\bk}, \lambda_{1\bk}$).  So now we can write the conduction
electron Green's function as:

\bea
\left[\mathcal{G}^c(\bk,i\omega)\right]^{-1}  = (i\omega_n - \epsilon_{0\bk}) - \epsilon_{1\bk} \tau_3 - \mathcal{V}\frac{i\omega_n - \lambda_{0\bk} + \lambda_{1\bk}\tau_3}{(i \omega_n - \lambda_{0\bk})^2-\lambda_{1\bk}^2} \mathcal{V}\dg
\eea
For completeness, the f-electron Green's function will be,
\bea
\left[\mathcal{G}^f(\bk,i\omega)\right]^{-1} = (i\omega_n - \lambda_{0\bk}) - \lambda_{1\bk} \tau_3 - \mathcal{V}\dg\frac{i\omega_n - \epsilon_{0\bk} + \epsilon_{1\bk}\tau_3}{(i \omega_n - \epsilon_{0\bk})^2-\epsilon_{1\bk}^2} \mathcal{V}.
\eea
The Green's function can be expanded in terms of $\vec{\tau}\otimes
\vec{\sigma}$, where $\vec{\sigma}$ describes the spin.  To do so, we
take $\tr \mathcal{V}\mathcal{V}\dg \tau_a \sigma_b$ and $\tr
\mathcal{V}\tau_3\mathcal{V}\dg \tau_a \sigma_b$.  The relevant
parameters are:
\bea 
V_{\bk+}^2 & =& \tr \mathcal{V}_\bk\mathcal{V}_\bk\dg = 2\tr
\left[\mathcal{V}_{6\bk} \mathcal{V}_{6\bk}\dg + \mathcal{V}_{7\bk}
\mathcal{V}_{7\bk}\dg\right]\cr
 V_{\bk-}^2 & =& 
\tr
\mathcal{V}_\bk\tau_3\mathcal{V}_\bk\dg\tau_3 = 2\tr
\left[\mathcal{V}_{6\bk} \mathcal{V}_{6\bk}\dg - \mathcal{V}_{7\bk}
\mathcal{V}_{7\bk}\dg\right]\cr 
\Delta_{\bk-} & = & \tr
\mathcal{V}_\bk\tau_3\mathcal{V}_\bk\dg\tau_2 = 2i \tr
\left[\mathcal{V}_{6\bk} \mathcal{V}_{7\bk}\dg - \mathcal{V}_{7\bk}
\mathcal{V}_{6\bk}\dg\right]\cr 
\vec{\Delta}_{\bk+} & = & \tr
\mathcal{V}_\bk\mathcal{V}_\bk\dg\tau_1 \vec{\sigma} = 2 \tr
\left[\left(\mathcal{V}_{6\bk} \mathcal{V}_{7\bk}\dg +
\mathcal{V}_{7\bk} \mathcal{V}_{6\bk}\dg\right)\vec{\sigma}\right].
\eea
All other combinations are uniformly zero.  The
conduction electron Green's function can be written in the
form $\left[\mathcal{G}^c\right]^{-1} = A \tau_0 + B\tau_3 +
\vec{C}\cdot \vec{\sigma}\tau_1 + D \tau_2$.  The eigenvalues are 
found by taking $\det \left[\mathcal{G}^c(\bk,\omega)\right]^{-1} =
0$, leading to an eighth order polynomial. For the
special particle-hole symmetric case where $\epsilon_0 = \lambda_0 =
0$, this reduces to $\omega^4 - 2 \alpha_\bk \omega^2
-\gamma_\bk^2$, which can be solved analytically. 
The four
(doubly degenerate) eigenvalues are $E_{\bk\eta}$ and are
found numerically on a grid of $\bk$ points.  Due to the
structure of the Green's function, we can write,
\begin{eqnarray}\label{l}
\mathcal{G}^c(i\omega_n, \bk) & = & \frac{1}{\prod_\eta (i\omega_n - E_{\bk \eta})} \left(A \tau_0 - B\tau_3 - \vec{C}\cdot \vec{\sigma}\tau_1 - D \tau_2\right),\cr
A & = & (i\omega_n-\epsilon_{0\bk})\left[(i \omega_n - \lambda_{0\bk})^2-\lambda_{1\bk}^2\right] - (i\omega_n -\lambda_{0\bk})V_{\bk +}^2\cr
B & = & -\epsilon_{1\bk}\left[(i \omega_n - \lambda_{0\bk})^2-\lambda_{1\bk}^2\right]-\lambda_{1\bk}V_{\bk -}^2\cr
\vec{C} & = & -(i\omega_n - \lambda_{0\bk})\vec{\Delta}_{\bk +}\cr
D & = & -\lambda_{1\bk} \Delta_{\bk -}
\end{eqnarray}
which 
makes it particularly easy to write down the conduction electron magnetization,
\bea
\vec{m}_c(\bQ)  &= & -(g \mu_B) T\sum_{\bk, \omega_{n}}\tr
\left[\vec{\sigma} \mathcal{G}^c(\bk,i\omega_n) \tau_1  \right]\cr
& &=  -(g \mu_B)\sum_{\bk \eta }\frac{(E_{\bk\eta} - \lambda_{0\bk})}{\prod_{\eta'\neq \eta} (E_{\bk\eta}- E_{\bk \eta'})}f(E_{\bk\eta})\vec{\Delta}_{\bk +}
\eea
and susceptibility, 
\bea
\chi^{xy} & = & -(g \mu_B)^2 T \sum_{i\omega_n}
\sum_{\bk}\tr\left[\sigma_x \mathcal{G}^c(\bk,i\omega_n)\sigma_y
\mathcal{G}^c(\bk,i\omega_n)\right]
\cr
& = & -(g \mu_B)^2\sum_{\bk \eta }\left[
 \frac{2 (E_{\bk \eta}-\lambda_{0\bk})f(E_{\bk\eta})+(E_{\bk \eta}-\lambda_{0\bk})^2f'(E_{\bk\eta})}{\prod_{\eta' \neq \eta} (E_{\bk\eta}-E_{\bk\eta'})^2}\right.\cr
&& \left.-\sum_{\eta'\neq \eta } \frac{2 (E_{\bk\eta}-\lambda_{0\bk})^2
f(E_{\bk\eta})}{
{(E_{\bk\eta}-E_{\bk\eta'})}\prod_{\eta'' \neq
\eta}(E_{\bk\eta}-E_{\bk\eta''})^2}\right]\Delta^{x}_{\bk +}\Delta^{y}_{\bk
+}
\eea
Note that above integral may be positive or negative.  The functions $f$ and $f'$ are the Fermi function, $f(x) = \left(\mathrm{e}^{-x/T}+1\right)^{-1}$ and it's derivative, $f'(x) = df(x)/dx$, respectively.

Although the conduction electrons develop a magnetic moment, the
f-electrons have no dipolar or quadrupolar moments, $\vec{m}_f = 0$.
Upon closer examination, the quadrupolar moments are found to vanish
because of a d-wave form factor, implying that there is no net
quadrupole moment and thus no associated lattice distortion, even for
uniform hastatic order. As a d-wave quadrupole is a hexadecapole,
this ultimately means that like Haule and Kotliar\cite{Haule09}, we
have staggered $(J_x J_y+J_y J_x)(J_x^2-J_y^2)$ hexadecapolar moments.
However, unlike Haule and Kotliar, where the hexadecapolar moments are
the primary order parameter (and thus of order one), here the
hexadecapolar moments are a seconday effect of the composite hastatic
order, and like the conduction electron moments, will be of order
$T_K/D$.  The f-electron moment has an identical form to the
conduction electron moment, (37), with $\epsilon \leftrightarrow
\lambda$ everywhere, and the relevant form-factor becomes
$\vec{\Delta}^f_{\bk +} = \tr \mathcal{V}_\bk \vec{\sigma}
\tau_1\mathcal{V}_\bk\dg$.  Given how difficult it is to observe large
hexadecapolar moments, the hexadecapolar moments associated with
hastatic order will almost certainly be unobservably small.  By
contrast, in the antiferromagnetic phase, the f-electrons develop a
large c-axis magnetic moment.

\section{Current neutron scattering bounds on the transverse moment}

We predict a staggered basal plane magnetic moment on the order of
$.01\mu_B$/U.  However, such a moment has not been previously
detected, despite a large number of neutron studies.  Early neutron
studies focused on the small, $.02 - .04 \mu_B$ c-axis
moment\cite{broholm91,walter93}, which was later shown to be due to
small, local regions of large moment
antiferromagnetism\cite{takagi07}.  Such inhomogeneous
antiferromagnetism persists even in high quality single crystals,
making it essential to identify the moment orientation.  The moment
orientation has not been examined since these early
works\cite{broholm91,walter93}, whose resolution of $.01 \mu_B$ is on
the order of our prediction.  As the neutron signal is proportional to
the component of the moment perpendicular to $\bQ$: ${\bf m}_\perp
\propto \bQ \times ({\bf m} \times \bQ)$, measuring along $\bQ =
(001)$ will isolate the transverse magnetic moment. More recent
studies have focused on $\bQ = (100)$, where the extrinsic c-axis
moment is always seen\cite{villaume}.  Detection of our predicted
${\bf m}_\perp$ will require high resolution, likely spin-polarized,
measurements along $\bQ = (001)$.

\section{Density of states}\label{}

To simplify the calculation of the ground-state properties, we have
taken
$\epsilon_{0\bk }= -\mu$ and $\lambda_{0\bk }= \lambda$ to be
constant, and chosen 
the filling of the conduction sea so that at $T=0$, 
$\lambda+\mu=0$.  In this case, the Hamiltonian ${\cal H} (\bk
)_{\alpha \beta }$ in (\ref{thematrix})
can be diagonalized
analytically, leading to four doublets 
$\vert  \bk \eta \sigma \rangle $
($\sigma  = \pm 1 $), 
with eigenvectors
\begin{equation}\label{}
u_{\alpha } (\bk \eta \sigma ) = \langle \alpha \vert \bk \eta  \sigma \rangle 
\end{equation}s
where ${\cal H}_{\alpha \beta } (\bk )\cdot u (\bk \eta \sigma )=
E_{\bk \eta }u (\bk \eta \sigma ),
$.  The degeneracy of the eigenvalues is guaranteed by the invariance of the Hamiltonian under time-reversal plus a translation by $\bQ$.
The energies $E_{\bk \eta }$ are given by 
\begin{equation}\label{}
E_{\bk \eta }  =\left\{ 
\lambda \pm \sqrt{\alpha_{\bk }\pm 
\sqrt{\alpha_{\bk}^{2}- \gamma_{\bk }^{2}}}\right\}
\end{equation}
where  $\alpha_{\bk }= V_{\bk +}^{2}+\frac{1}{2}\left(\epsilon_{1\bk
}^{2}+\lambda_{1\bk }^{2} \right)$ and 
$\gamma_{\bk }^{2}= (\epsilon_{1\bk }\lambda_{1\bk } - V_{\bk
-}^{2})^{2} + (\vec{\Delta}_{\bk -})^{2}$.

The total density of states is then given by 
\begin{eqnarray}\label{l}
A (\omega) = \sum_{\bk \eta } \delta (\omega- E_{\bk
\eta })= \frac{1}{\pi}{\rm Im}\int \frac{d^{3}k}{(2\pi)^{3}} \sum_{  \eta}
\frac{1}{\omega - E_{\bk \eta }-i \delta 
}
\end{eqnarray}
where the integral is over the Brillouin zone.
Numerically, this quantity was computed by a sum over a discrete
set of momenta, dividing the Brillouin zone into $40^{3}$ points and
using a small value of $\delta $ to broaden the delta-function into a Lorentzian.

\section{g-factors}\label{}

The Zeeman energy is determined by the Hamiltonian 
\begin{equation}\label{}
- \vec{B}\cdot\vec{M} = -\sum_{\bk  \in \frac{1}{2}BZ} \psi \dg_{\bk }{\vec{\cal M}}
\psi_{\bk }\cdot \vec{B}
\end{equation}
where $\psi_{\bk }= (c_{\bk }, c_{\bk +\bQ }, \chi _{\bk },\chi_{\bk
+\bQ } )^{T}$ and 
\begin{equation}\label{}
\vec{{\cal M}}= \frac{1}{2}\left(\begin{matrix} 
2 \mu_B \vec{\sigma}
& 0  & 0 & 0 \cr
0 & 2 \mu_B \vec{\sigma}
 & 0  & 0 \cr
0 & 0 & {g_{f}} \mu_B {\sigma^{z}} & 0 \cr
0 & 0 & 0 & {g_{f}}\mu_B {\sigma^{z}}
\end{matrix}
 \right), 
\end{equation}
where 
$g_{f}$ is the effective g-factor of the Ising Kramers doublet.
In a field, the doubly-degenerate energies, $\vert  \bk \eta \sigma \rangle $
($\sigma = \pm 1$) are split apart so that 
$\Delta E_{\bk \eta }= |E_{\bk \eta \up}-E_{\bk \eta \dw }|=  g_{\bk \eta
} (\theta ) B$, 
so the g-factor is given by  $g_{\bk \eta } (\theta ) = \left \vert \frac{d\Delta E_{\bk \eta }}{dB}\right\vert_{B\rightarrow 0}
$.
Now we are interested in the Fermi surface average of the g-factor,
given by 
\[
g (\theta ) = \frac{
\sum_{\bk \eta } g_{\bk \eta } (\theta )\delta (E_{\bk\eta })
}{\sum_{\bk \eta } \delta (E_{\bk\eta })
}
\]
These quantities were 
calculated numerically, on a $40^{3}$ grid, 
using $g_{f}=2.9$ for 
the effective g-factor of the local Kramers doublet. The resulting 
g-factor in the $z-$ direction is reduced to $g_{eff} (\theta =0)=
2.6$ because of the admixture with conduction electrons. 
The 
delta-functions were treated as narrow Lorentzians $\delta (E)= \frac{1}{\pi }Im
(E-i\eta )^{-1}$, where $\eta $ is a small postive number. 
The g-factors at each point in momentum space were computed
by introducing a small field $\delta B$
into the Hamiltonian, with the approximation $g_{\bk \eta } (\theta ) = |E_{\bk \up} -E_{\bk \dw} |/\delta B$.

\section{Model Tunneling conductance and nematicity calculation}

To calculate the tunneling density of states, we assume that the
differential conductance is proportional to the local Green's function
on the surface of the material
\begin{equation}\label{l}
\frac{dI}{dV} (\bx ) \propto  A (\bx, eV)
\end{equation}
where
\[
A (\bx ,\omega ) = \frac{1}{\pi} {\rm Im }G_{\sigma \sigma } (x, \omega- i\delta )
=\sum_{\sigma } \int_{-\infty }^{\infty }dt  \langle \{
\psi_{\sigma } (\bx ,t), \psi_{\sigma } \dg(\bx,0)\}\rangle  e^{i \omega t}
\]
is the imaginary part of the local electronic Green's function.
To calculate this quantity, we  decompose the local electron
field in terms of the low energy fermion modes of the system, writing
\begin{equation}\label{}
\psi _{\sigma } (\bx ) = \sum_{j}\left[
\phi_c (|\bx - \bR_{j}|)
c_{j\sigma } +
\phi^{6}_{\sigma \alpha } (\bx - \bR_{j}) f_{j\Gamma_{6}\alpha}
+\phi^{7}_{\sigma \alpha }  (\bx - \bR_{j})f_{j\Gamma_{7}\alpha} \right]
\end{equation}
where $\phi_{c} (|\bx -\bR_{j}|)$ is the wavefunction of the conduction electron
centered at site j, while
\begin{eqnarray}\label{l}
\phi^{6}_{\sigma \alpha } &=& \phi^{6} (|\bx  - \bR_{j}|){\cal
Y}^{6}_{\sigma \alpha } (\bx  - \bR_{j})\cr
\phi^{7}_{\sigma \alpha } &=& \phi^{7} (|\bx  - \bR_{j}|){\cal
Y}^{7}_{\sigma \alpha } (\bx  - \bR_{j})
\end{eqnarray}
are the wave functions of the $\Gamma_{6} $ and $\Gamma_{7^{-}}$ 
f-orbitals  centered at site $j$.

Projected into the low energy subspace, we have 
$f_{j\Gamma_{6}\alpha } \rightarrow (\langle B\dg_{j}\rangle
\chi_{j})_{\alpha } $ and 
$f_{j\Gamma_{7}\alpha } \rightarrow (\langle B\dg_{j}\rangle \sigma_{1}\chi_{j})_{\alpha }$.
Writing $B\dg _{j} = b U_{j}$, and $\tilde{\chi }_{j}= U_{j}\chi_{j}$
these expressions become 
$f_{j\Gamma_{6}\alpha } \rightarrow b \tilde{ \chi }_{j}$ and 
$f_{j\Gamma_{7}\alpha } \rightarrow b (\hat {\bf n}\cdot \vec{\sigma
})e^{-i \bQ  \cdot \bR_{j}}\tilde\chi_j$. Making these substitutions 
and rewriting the field operators in momentum space we obtain
\begin{eqnarray}\label{l}
\psi _{\sigma } (\bx )  
= \sum_{j\bk \in \frac{1}{2}BZ}
e^{-i\bk \cdot \bR_{j}}
\Lambda_{j}(\bx - \bR_{j})
\cdot \left(
\begin{matrix} c_{\bk \alpha }\cr c_{\bk+\bQ\alpha}\cr
\chi_{\bk \alpha }\cr
\chi_{\bk+\bQ \alpha }
\end{matrix} \right)
\end{eqnarray}
where
\begin{eqnarray}\label{l}
\Lambda_{j} (\bx ) &=&
\left( \right.
\phi_c (|\bx |)\delta_{\sigma \alpha },
e^{-i \bQ \cdot \bR_{j}}
\phi_c (|\bx |)\delta_{\sigma \alpha },\cr
&& \left. b\phi^{6}_{\sigma
\alpha } (\bx ) + e^{-i \bQ \cdot \bR_{j}}
b\phi^{7}_{\sigma \alpha }  (\bx ) (\hat  {\bf n}\cdot \vec{\sigma })
,
 e^{-i \bQ \cdot \bR_{j}}b\phi^{6}_{\sigma
\alpha } (\bx ) +
b\phi^{7}_{\sigma \alpha }  (\bx ) (\hat  {\bf n}\cdot \vec{\sigma })
 \right)
\end{eqnarray}

We choose a layer where $e^{-i(\bQ\cdot\bR_{j})}=+1$, then on this layer
the local Green's function is given by 
\begin{eqnarray}\label{l}
G (\bx ,\omega)=
\sum_{j,l}
\tilde{ \Lambda} (\bx
 - \bR_{j})\cdot {\cal G}_{jl} (\omega)\cdot \tilde{\Lambda}\dg (\bx -\bR_{l}) 
\end{eqnarray}
where 
\[
{\cal G}_{jl} (\omega)= \sum_{\bk  \in \frac{1}{2 }BZ}{\rm Tr}[(1+ \tau_{1}) {\cal G} (\bk ,\omega)] e^{-i \bk \cdot(\bR_j-
\bR_{l})}
\]
is a trace only over the momentum degrees of freedom, so ${\cal G}_{jl}$ is a
four by four matrix for each pair of lattice points $j$ and $l$, where
$\tilde{ \Lambda} (\bx) = 
\left( 
\phi_c (|\bx |)\delta_{\sigma \alpha },
b\phi^{6}_{\sigma
\alpha } (\bx ) 
+ b\phi^{7}_{\sigma \alpha }  (\bx ) 
(\hat  {\bf n}\cdot \vec{\sigma })
 \right)$. 

The final spectral function is then 
\begin{equation}\label{}
A (\bx ,\omega ) = \frac{1}{\pi} {\rm Im \ Tr}\left[
\sum_{j,l} \tilde{ \Lambda} (\bx - \bR_{j})\cdot {\cal G}_{jl} (\omega-i\delta )\cdot \tilde{\Lambda}\dg (\bx -\bR_{l}) 
 \right]
\end{equation}
To evaluate this quantity,  the summations were limited to the four
nearest neighbor sites at the corner of a plaquette.  The positions $\bx
$ were taken to lie in the plane of the $U$ atoms.
The wavefunctions 
$\phi^{6} (\vert \bx \vert )= e^{-\vert  \bx  \vert /a}$, $\phi^{7}
(\vert \bx \vert )= e^{-\vert  \bx  \vert /a}$ and 
$\phi_{c} (\vert \bx \vert )= e^{-\vert  \bx  \vert /a}$ were each
taken to be simple exponentials of characteristic range equal to the
$U-U$ spacing a.

The nematicity of the tunneling conductance was then calculated
numerically from the spatial integral
\begin{equation}
\eta (eV) = 
\frac{
\int A (\bx ,eV)
\ {\rm sgn} (xy)dxdy
}
{\left(
\int dx dy A (\bx ,eV)^{2}
- 
\left[ \int dx dy A (\bx ,eV)\right]^{2}
\right)^{1/2}
}
\end{equation}